\documentclass[a4paper, 11pt]{article}
\pdfoutput=1

\usepackage{jinstpub} 

\title{Commissioning of the vacuum system of the KATRIN Main Spectrometer}

\newcommand{\bonn}{a}
\newcommand{\ekp}{b}
\newcommand{\ucsb}{c}
\newcommand{\massit}{d}
\newcommand{\muenster}{e}
\newcommand{\mainz}{f}
\newcommand{\ipe}{g}
\newcommand{\itep}{h}
\newcommand{\ikp}{i}
\newcommand{\uw}{j}
\newcommand{\ppq}{k}
\newcommand{\unc}{l}
\newcommand{\npi}{m}
\newcommand{\wuppertal}{n}
\newcommand{\madrid}{o}
\newcommand{\swansea}{p}
\newcommand{\fulda}{q}
\newcommand{\inr}{r}
\newcommand{\lbnl}{s}

\newcommand{\bonnAddress}{Helmholtz-Institut fuer Strahlen- und Kernphysik, University Bonn, Nussallee 14-16, 53115 Bonn, Germany}
\newcommand{\fuldaAddress}{University of Applied Sciences (FH) Fulda, Leipziger Str.~123, 36037 Fulda, Germany}
\newcommand{\ekpAddress}{Institute of Experimental Nuclear Physics (IEKP), Karlsruhe Institute of Technology (KIT), Wolfgang-Gaede-Str. 1, 76131 Karlsruhe, Germany}
\newcommand{\ikpAddress}{Institute for Nuclear Physics (IKP), Karlsruhe Institute of Technology (KIT), Hermann-von-Helmholtz-Platz 1, 76344 Eggenstein-Leopoldshafen, Germany}
\newcommand{\ipeAddress}{Institute for Data Processing and Electronics (IPE), Karlsruhe Institute of Technology, (KIT)Hermann- von-Helmholtz-Platz 1, 76344 Eggenstein-Leopoldshafen, Germany}
\newcommand{\itepAddress}{Institute for Technical Physics (ITeP), Karlsruhe Institute of Technology (KIT), Hermann- von-Helmholtz-Platz 1, 76344 Eggenstein-Leopoldshafen, Germany}
\newcommand{\ppqAddress}{Project, Process, and Quality Management (PPQ), Karlsruhe Institute of Technology (KIT), Hermann- von-Helmholtz-Platz 1, 76344 Eggenstein-Leopoldshafen, Germany    }
\newcommand{\inrAddress}{Academy of Sciences of Russia, Institute for Nuclear Research, 60th October Anniversary, Prospect 7a, 117312 Moscow, Russia}
\newcommand{\lbnlAddress}{Institute for Nuclear and Particle Astrophysics and Nuclear Science Division, Lawrence Berkeley National Laboratory, Berkeley, CA 94720, USA}
\newcommand{\madridAddress}{Universidad Complutense de Madrid, Instituto Pluridisciplinar, Paseo Juan XXIII, n\textsuperscript{\b{o}} 1, 28040 - Madrid, Spain}
\newcommand{\mainzAddress}{Institut f\"{u}r Physik, Johannes-Gutenberg-Universit\"{a}t Mainz, 55099 Mainz, Germany}
\newcommand{\massitAddress}{Laboratory for Nuclear Science, Massachusetts Institute of Technology, 77 Massachusetts Ave, Cambridge, MA 02139, USA}
\newcommand{\muensterAddress}{Institut f\"{u}r Kernphysik, Westf\"{a}lische Wilhelms-Universit\"{a}t M\"{u}nster, Wilhelm-Klemm-Str. 9, 48149 M\"{u}nster, Germany}
\newcommand{\npiAddress}{Nuclear Physics Institute of the CAS, v. v. i., CZ-250 68 \v{R}e\v{z}, Czech Republic}
\newcommand{\swanseaAddress}{Department of Physics, Swansea University, Singleton Park, Swansea SA2 8PP, United Kingdom}
\newcommand{\ucsbAddress}{Department of Physics, University of California at Santa Barbara, Santa Barbara, CA 93106, USA}
\newcommand{\uncAddress}{Department of Physics and Astronomy, University of North Carolina, Chapel Hill, NC 27599, USA}
\newcommand{\uwAddress}{Center for Experimental Nuclear Physics and Astrophysics, and Dept.~of Physics, University of Washington, Seattle, WA 98195, USA}
\newcommand{\wuppertalAddress}{Department of Physics, Faculty of Mathematics und Natural Sciences, University of Wuppertal, Gauss-Str. 20, D-42119 Wuppertal, Germany}

\affiliation[\bonn]{\bonnAddress}
\affiliation[\ekp]{\ekpAddress}
\affiliation[\ucsb]{\ucsbAddress}
\affiliation[\massit]{\massitAddress}
\affiliation[\muenster]{\muensterAddress}
\affiliation[\mainz]{\mainzAddress}
\affiliation[\ipe]{\ipeAddress}
\affiliation[\itep]{\itepAddress}
\affiliation[\ikp]{\ikpAddress}
\affiliation[\uw]{\uwAddress}
\affiliation[\ppq]{\ppqAddress}
\affiliation[\unc]{\uncAddress}
\affiliation[\npi]{\npiAddress}
\affiliation[\wuppertal]{\wuppertalAddress}
\affiliation[\madrid]{\madridAddress}
\affiliation[\swansea]{\swanseaAddress}
\affiliation[\fulda]{\fuldaAddress}
\affiliation[\inr]{\inrAddress}
\affiliation[\lbnl]{\lbnlAddress}

\author[\bonn]{M.~Arenz,}
\author[\ekp]{M.~Babutzka,}
\author[\ucsb]{M.~Bahr,}
\author[\massit]{J.P.~Barrett,}
\author[\muenster]{S.~Bauer,}
\author[\mainz]{M.~Beck,}
\author[\ipe]{A.~Beglarian,}
\author[\muenster]{J.~Behrens,}
\author[\ipe]{T.~Bergmann,}
\author[\itep]{U.~Besserer,}
% ...

\emailAdd{joachim.wolf@kit.edu}

\collaboration{KATRIN Collaboration}

\abstract{The KATRIN experiment will probe the neutrino mass by measuring
the $\beta$-electron energy spectrum near the endpoint of tritium $\beta$-decay.  
An integral energy analysis will be performed by an electro-static spectrometer (``Main Spectrometer''), 
an ultra-high vacuum vessel with a length of 23.2$\,$m, a volume of 1240$\,$m$^3$, and a 
complex inner electrode system with about 120\,000 individual parts. 
The strong magnetic field that guides the $\beta$-electrons is provided by  
super-conducting solenoids at both ends of the spectrometer. Its influence on turbo-molecular 
pumps and vacuum gauges had to be considered. 
A system consisting of 6 turbo-molecular pumps and 3~km of non-evaporable getter strips 
has been deployed and was tested during the commissioning of the spectrometer. In this paper the 
configuration, the commissioning with bake-out at $300\,^\circ$C, and the performance of this system are 
presented in detail. The vacuum system has to maintain a pressure in the $10^{-11}\,$mbar range. 
It is demonstrated that the performance of the system is already close to these stringent functional requirements 
for the KATRIN experiment, which will start at the end of 2016. }


\begin{document}

\maketitle
\flushbottom

\section{Introduction \label{sec:intro}}

The {\bf KA}rlsruhe {\bf TRI}tium {\bf N}eutrino experiment (KATRIN) is
designed to determine the effective mass of electron anti-neutrinos with an
unprecedented sensitivity of $0.2\,$eV/c$^2$ at 90\% CL. This will be accomplished 
by measuring the shape of the energy spectrum of electrons from tritium $\beta$-decay
\cite{lit:Dsgn04,lit:Overview13}. The analysis is focused on
the last few eV below the 18.6\,keV endpoint of the $\beta$-spectrum.
An integrating, electrostatic spectrometer 
of MAC-E-Filter\footnote{Magnetic Adiabatic Collimation combined with an Electrostatic Filter} type
\cite{lit:Lobashev,lit:Picard} can provide high energy resolution with a  
wide open solid angle acceptance for $\beta$-electrons,
emitted isotropically in the tritium source. This technique has been successfully employed with
different types of tritium sources in the Mainz and Troitsk experiments, which
provide the most stringent, model-independent limits on the effective neutrino mass
\cite{lit:Mainz,lit:Troitsk,lit:PDG2006}: 
\begin{equation} 
m(\bar{\nu}_{\rm e}) = \sqrt{\sum_{i=1}^{3}|U_{{\rm e}i}^2|\cdot m_i^2} < 2\,{\rm eV/c^2 \quad (95\% C.L.}) .
\end{equation}
The effective mass $m(\bar{\nu}_e)$  is the incoherent sum of the three neutrino mass eigenstates $m_i$,
weighted with the mixing matrix coefficients $U_{{\rm e}i}$ \cite{lit:otten2008}.

\begin{figure} \centering
\includegraphics[width = \textwidth]{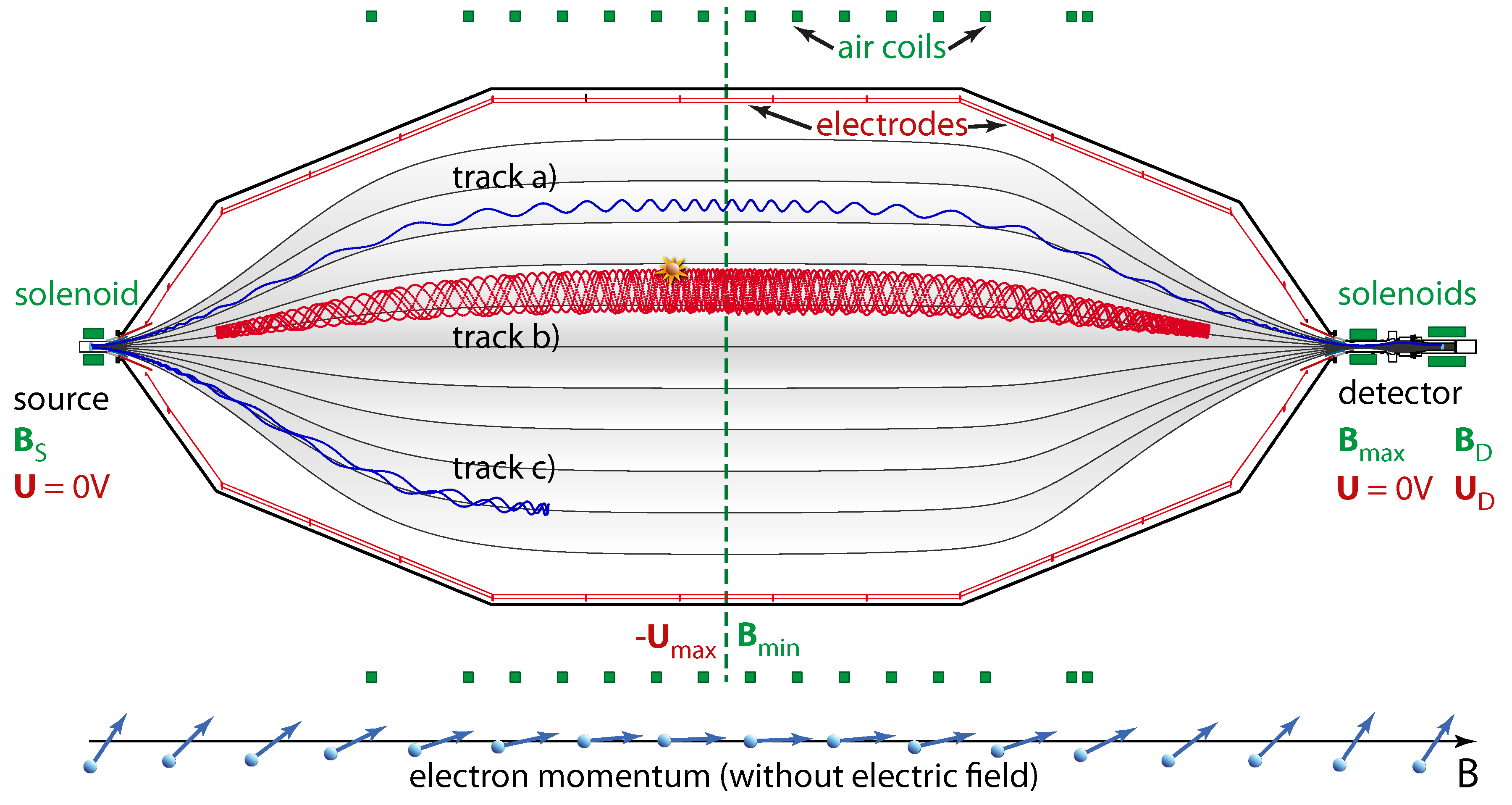} 
\caption{The principle of a MAC-E-Filter. The shaded area marks the magnetic flux tube connecting the source with the detector. The dashed line at the center indicates the ``analyzing plane'' of the MAC-E-Filter, where the magnetic field is at its minimum $B_{\rm min}$ and the electrostatic potential barrier at its maximum ($-U_{\rm max}$).  Electrons, originating from the source, are magnetically guided against the electrostatic retarding field towards the detector. Track a) is the trajectory of an electron with enough kinetic energy to overcome the retarding potential (cyclotron radius not to scale). The electron following track c) has less energy and is reflected back to the source. Track b) belongs to a magnetically trapped electron that has been created inside the MAC-E-Filter, for instance by a radioactive decay. The arrows at the bottom indicate the direction of the momentum of an electron relative to the guiding magnetic field line. The inhomogeneous field transforms transverse momentum into longitudinal momentum and back.} \label{fig:MAC-E-Filter}
\end{figure}

The main features of a MAC-E-filter are illustrated in Fig.~\ref{fig:MAC-E-Filter}.
The $\beta$-electrons are adiabatically guided by strong 
magnetic fields from their point of origin in the tritium source ($B_{\rm S}$) through the MAC-E-filter. 
The superconducting solenoids at both ends provide the magnetic guiding field.  
For the KATRIN Main Spectrometer additional air coils induce a weak 
guiding field at the center of the spectrometer, and compensate for distortions by the 
earth magnetic field, as well as fringe fields of the solenoids and residual magnetization. 
The electrons move along the field lines in cyclotron motions, allowing for an accepted solid angle 
of almost $2\pi$. On their way to the central plane (``analyzing plane'') of the spectrometer, 
the magnetic field drops by several orders of magnitude to $B_{\rm min}$.   
Due to the slowly varying field, the transverse momentum of the cyclotron motion is adiabatically 
transformed into longitudinal momentum, parallel to the field lines. 
In short, the \emph{magnetic adiabatic transformation} transforms the isotropically emitted 
$\beta$-electrons at the source into a broad, parallel beam of electrons at the center of the 
MAC-E-filter. 

With both ends at ground potential and a high negative electric potential at its center ($-U_{\rm max}$), the MAC-E filter 
works as an electrostatic high-pass energy filter, reflecting all electrons with energies below the retarding 
potential (Fig. 1, track c). All other electrons are accelerated again towards the far end of the spectrometer, where they 
are counted by a detector (Fig. 1, track a). An optional positive potential at the detector ($U_{\rm D}$) can 
accelerate the electrons further, shifting their energy further away from the low energy ambient background radiation. 
The energy spectrum is measured by varying the retarding voltage around the 
endpoint energy of the $\beta$-spectrum. The energy resolution of the MAC-E-filter is limited by the remaining 
transverse energy of the cyclotron motion, which cannot be analyzed with the retarding potential. Assuming 
the conservation of the magnetic moment of the cyclotron motion, the energy resolution is defined by the 
ratio of the weak magnetic field $B_{\rm min}$ at the analyzing plane and the strongest magnetic field 
$B_{\rm max}$ along the trajectory of an electron with energy $E_{\rm e}$ \cite{lit:Picard}:
\begin{equation} 
\Delta E = \frac{B_{\rm min}}{B_{\rm max}}\cdot E_{\rm e}. \label{Equ:deltaE}
\end{equation} 
With a count rate of $10^{-2}\,$counts per second (cps) in the last eV below the endpoint of the 
$\beta$-spectrum, the KATRIN experiment requires not only higher statistics and improved energy 
resolution, but also aims for a low total background rate of similar size, 
in order to achieve an order-of-magnitude improvement in $m(\bar{\nu}_{\rm e})$ sensitivity.

A major source of background can arise from keV-range electrons originating from the radioactive
decays of neutral atoms or molecules inside the spectrometer volume, such as radon
\cite{lit:RadonFF,lit:RadonNW} 
and tritium \cite{lit:RadonSM}. 
If the decays occur within the magnetic flux tube inside the spectrometer, 
many of these primary electrons can be trapped by the magnetic mirror effect (Fig. 1, track b). 
The trapped electrons circulate inside the spectrometer for hours, until they have 
lost enough energy through ionization of residual gas molecules to leave the trap.
The low-energy secondary electrons can leave the trap through either end of the spectrometer. 
Being accelerated by the retarding potential of the spectrometer, about half of them can reach the 
detector with exactly the same energy as the signal electrons from tritium decay, thereby 
increasing the background rate.  
The number of secondary electrons produced depends on the energy of the primary electron. The storage time,
and thus the background rate, depend on the pressure in the spectrometer volume. 
Therefore the vacuum system of the KATRIN experiment \cite{lit:JVS09} is a key component for 
reducing this kind of background. Most of the few tritium molecules that reach the 
spectrometer can be pumped out before their decay (half-life: 12.1\,years). 
Short-lived isotopes, such as $^{219}$Rn with a 
half-life of 4\,s, are more likely to decay inside the spectrometer. A pressure of $10^{-11}\,$mbar is 
needed in order to extend the storage time long enough for removing the primary background 
electrons by active methods, such as electric or magnetic pulsing \cite{lit:ECR_SM}, 
before they can produce too many secondary electrons. 

The subjects of this paper are the commissioning of the complete Main
Spectrometer vacuum system in the first half of 2013, the conditioning of the large vessel by
vacuum-baking, various remedies for technical problems that were encountered
and the vacuum performance during
the first electron measurements with the MAC-E-filter system. The next section
provides a short overview of the KATRIN experiment. Section~\ref{sec:vacuum}
describes the vacuum system of the Main Spectrometer.
In Section~\ref{sec:simulation} the vacuum simulations, needed for the interpretation of the measured 
data, are described. Section~\ref{sec:commissioning} gives an account of the bake-out procedure of the
spectrometer and explains the methods used to quantify the performance of the
vacuum system. It is followed by section~\ref{sec:argon}, describing the solution for the problem 
of a defective valve, which led to the venting of the whole spectrometer with
ultra-clean argon to prevent deactivating the NEG pumps. Finally we draw some
conclusions on the lessons learned in Section~\ref{sec:conclusions}.


\section{The KATRIN experiment \label{sec:katrin}}

\begin{figure} \centering
\includegraphics[width = \textwidth]{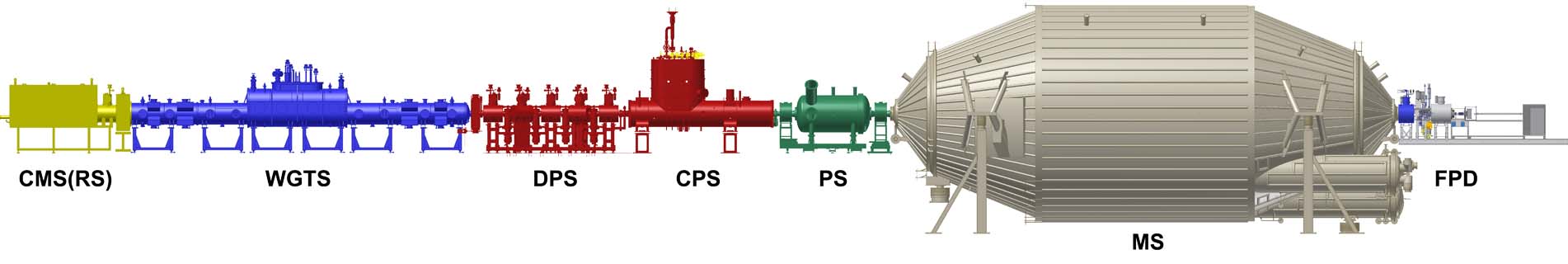} 
\caption{Overview of the 70$\,$m long
KATRIN experiment: calibration and monitoring system (CMS), windowless gaseous
tritium source (WGTS), differential pumping section (DPS), cryogenic pumping
section (CPS), Pre-Spectrometer (PS), Main Spectrometer (MS), focal plane detector (FPD).
For better clarity the Main Spectrometer is shown without the surounding magnetic air coil system.}
\label{fig:KATRIN}
\end{figure}

The main components of the KATRIN experiment \cite{lit:Dsgn04} are shown in
Fig~\ref{fig:KATRIN}. The 70$\,$m long system, currently under construction at
the Karlsruhe Institute of Technology (KIT), can be subdivided into the
\emph{Source \& Transport Section} (STS), where the tritium decays take place, and 
the \emph{Spectrometer \& Detector Section} 
(SDS), where the energies of the decay electrons are measured.

\subsection{The source and transport section \label{sec:STS}}

The STS has four main components. The central part is the \emph{Windowless
Gaseous Tritium Source} (WGTS), where molecular tritium gas is injected at the
center of a 10$\,$m long, 90$\,$mm diameter tube, and where most of the $\beta$-decays take 
place. The $\beta$-activity inside the source tube will be around $10^{11}$\,Bq. 
The tube is differentially pumped at both ends by turbo-molecular pumps (TMP),
which remove 99\% of the gas. The tritium is recirculated through a closed loop system
\cite{lit:WGTSmon,lit:InnerLoop}. At the rear end of the WGTS, the \emph{Calibration and
Monitoring System} (CMS) measures the tritium activity by monitoring the flux of
incoming $\beta$-electrons. In addition it provides mono-energetic electrons from an
electron source with well-defined energy and emission angle for the calibration of the
experiment \cite{lit:rear_egun}. Between the WGTS and the spectrometer section
two additional pumping systems remove most of the remaining tritium gas,
reducing the total flux by a factor of more than $10^{14}$. The first stage is
again a \emph{Differential Pumping Section} (DPS), using TMPs \cite{lit:DPS}.
The second stage is a \emph{Cryogenic Pumping Section} (CPS), where cryosorption
on  argon frost at $3\,$K is used to capture tritium molecules \cite{lit:CPS1,lit:CPS2}.
Throughout the STS superconducting solenoids produce magnetic fields between 0.5
and $5.6\,$T, guiding around $10^{10}$ $\beta$-electrons per second 
adiabatically to the spectrometer and detector section.

\subsection{The spectrometer and detector section \label{sec:SDS}}

The SDS consists of three main components, the \emph{Pre-Spectrometer}
(PS) with a moderate energy resolution of $70\,$eV, followed by the \emph{Main
Spectrometer} (MS), where the energy of electrons is analyzed with a resolution
of $0.93\,$eV, and the \emph{Focal Plane Detector} (FPD), which counts electrons 
that have passed the retarding voltages of both MAC-E-Filters.

The PS serves several purposes. It can work as a pre-filter, rejecting all electrons
with energies more than 300\,eV below the endpoint energy of the
$\beta$-spectrum, thus reducing the electron flux into the MS by seven orders of
magnitude to about $10^3$ electrons per second. The magnetic guiding field of the MAC-E-filter is induced by
4.5\,T superconducting solenoids at both ends of the spectrometer. 
The vacuum vessel of the PS, with a diameter of 1.7$\,$m and a length of
3.4$\,$m, served as a prototype for the vacuum system of the MS. The vacuum
system of the PS uses a combination of \emph{non-evaporable-getter} (NEG) pumps
made of 90$\,$m of 30$\,$mm wide SAES St707\textsuperscript{\textregistered} NEG strips and two TMPs
\cite{lit:PS_outgassing}, providing a base pressure of $10^{-11}\,$mbar. 
The vacuum pumps also reduce the small incoming flux of tritiated
molecules from the STS to the MS by another two orders of magnitude.
Although the NEG strips have been identified as a
major source of radon-related background \cite{lit:RadonFF}, there is no
alternative pumping concept with which to obtain the huge pumping speed needed to operate the MS,
since helium-cooled cryogenic pumps have much higher operating costs. However,
the tests with the PS showed that LN$_2$ cooled baffles in front of the getter pumps
are able to suppress the Rn-induced background effectively.  
  
The high energy resolution of the MS of 0.93\,eV at 18.6\,keV requires a ratio between 
$B_{\rm min}$ and $B_{\rm max}$ (see Equ.~\ref{Equ:deltaE} and Fig.~\ref{fig:MAC-E-Filter})
of 1:20\,000.  The magnetic guiding field is generated by two 
superconducting solenoids at the detector side at 6\,T and 3.5\,T, respectively. At the other end a
4.5$\,$T solenoid is shared between the MS and the PS. In addition the MS is surrounded by a 
system of air-coils with
a diameter of 12.6$\,$m. It compensates for the earth magnetic field and solenoid fringe fields, and confines
the flux-tube of the magnetic guiding field inside the volume of the MS
\cite{lit:air-coil}. Together these components generate a field layout with a very 
high degree of axial symmetry, which provides magnetic shielding for low energy electrons
emitted from the spectrometer walls, reducing cosmic-ray-induced background
by a factor of $10^{5}$ \cite{lit:phd:leiber}.

\begin{figure} 
\centering
\includegraphics[width = \textwidth]{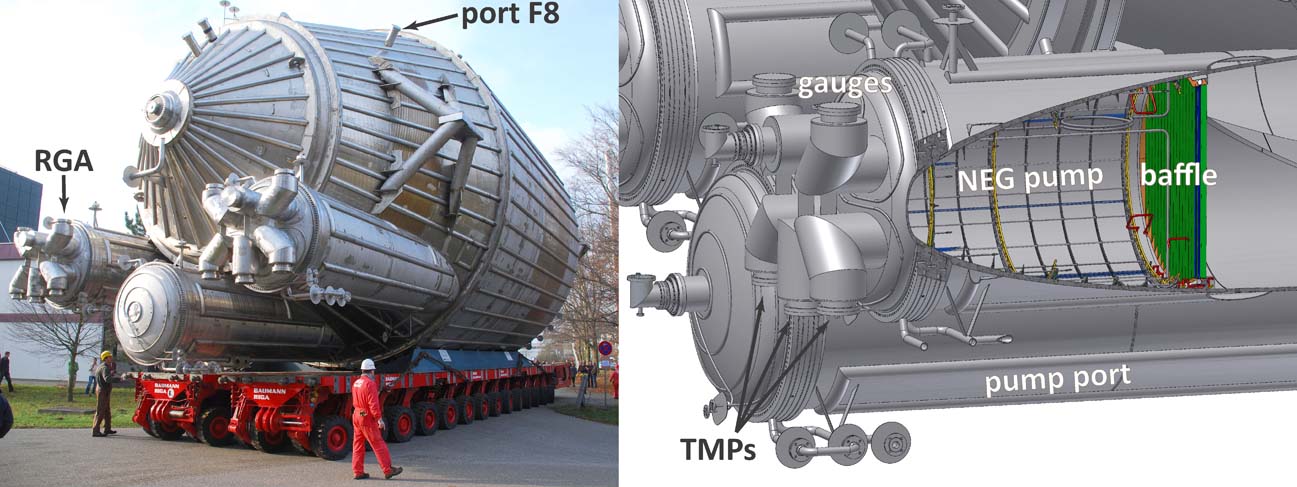} 
\caption{Left: Arrival of the KATRIN Main
Spectrometer vacuum vessel at the Karlsruhe Institute of Technology. 
One of the 50-cm-long DN200 ports is indicated. 
Right: Location of the main vacuum pumps in one of the three pump ports.} \label{fig:mainspec} 
\end{figure}

The requirement of adiabatic electron transport with a slowly varying B-field in the MAC-E filter, 
and the cross section of the magnetic flux-tube at the analyzing plane, which scales inversely 
with the field strength, imply a very large MS (see Fig.~\ref{fig:mainspec}). 
The stainless steel (316LN) vacuum vessel has a total
length of $23.2\,$m, a diameter of $9.8\,$m and a weight of 200$\,$t.
 
The vacuum pipes of the electron beam-lines at both ends of the
spectrometer terminate in two axisymmetric aluminum cones, which are held 
at ground potential as anodes. They are connected to the vacuum vessel
via ceramic insulators. The electrostatic retarding field of the MAC-E filter is generated by
connecting the outer hull of the MS to a high precision high voltage system
($-18.5\,$kV), which has to be stabilized and monitored with parts-per-million (ppm) accuracy
\cite{lit:MS_HVdivider}. Together with a complex wire electrode system that is mounted to the inner wall,
the vessel acts as the cathode. Between the ground electrode and the first and the last
ring of the wire electrodes, conical electrodes that are formed from titanium sheet metal are 
maintained at the vessel potential. These so-called anti-Penning electrodes act as 
shielding in the high-field region to prevent deep Penning traps from forming. 

\begin{figure} 
\centering
\includegraphics[width = \textwidth]{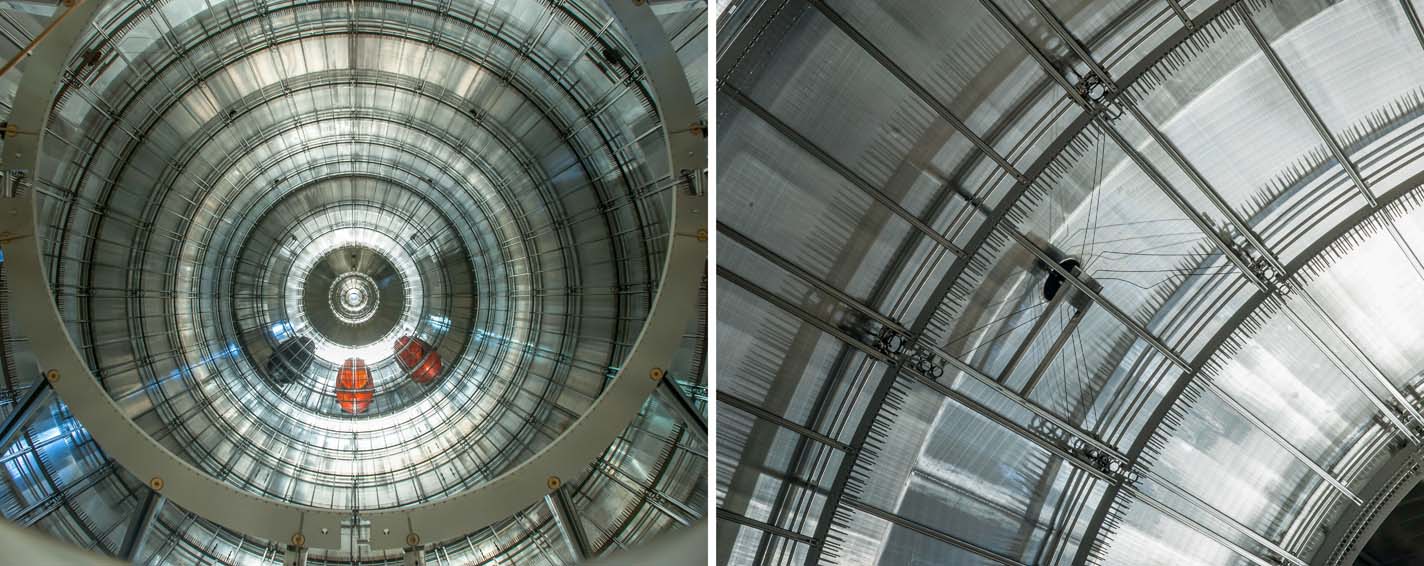} 
\caption{Left: View of the complete wire-electrode system as seen from the beam-line flange at the source 
end of the MS. At the far end the three pump ports with the LN$_2$ baffles are visible. Right: Voltage distribution 
to the corners of an electrode module from the distribution panel 
underneath a flange with electrical feedthroughs.} \label{fig:electrodes} 
\end{figure}

The wire-electrode system consists of 23,440 individually insulated wires (see Fig.~\ref{fig:electrodes}).  
It is used for fine-tuning
the electrostatic field, preventing Penning traps, and providing the axial symmetry of the field \cite{lit:MSelectrode}.
With the wires being at a potential that is 100$\,$V lower than the vessel, 
the system is also responsible for the electrostatic
rejection of electrons created by cosmic muons or radioactive decays at the wall of the
vessel. The wires are strung on 248
stainless steel frames (``modules''). In most of these electrode modules the wires are
strung in two layers. In addition the electrode system is subdivided both in the axial 
direction and in the vertical direction into several sections.
This allows for a gradual adjustment of the electric potential in the axial direction, and for 
applying short dipole pulses regularly to remove magnetically trapped electrons from the MS.  
Modules belonging to the same section share the same voltages for their wire layers. 
Each section contains between 4 and 50 modules.  

The high voltage vacuum feedthroughs are mounted at DN200 ports above the different sections.
Inside the vacuum volume, the feedthroughs are connected with 1.5-mm diameter stainless steel
 (Inconel\textsuperscript{\textregistered}) 
wires to the insulated connectors at the distribution panels that are attached to the frames of the electrode modules 
underneath the respective ports.  Copper-beryllium (CuBe) rods with a diameter of 3$\,$mm 
distribute the voltages from the distribution panels to the corners of the first module of a section,
where further distributions to neighboring modules are achieved via spring-loaded contacts and short wires. 

Short circuits between wire
layers would reduce the efficiency of background rejection, while a broken wire,
which may electrically short to the vessel, would render both the fine tuning 
of the field and the rejection of backgrounds ineffective. Special care and extensive quality
control measures were taken to build a robust wire-electrode system, in particular with
regard to the stress on the numerous wires and interconnects during the 
bake-out of the vacuum system. 

When the electrons leave the MS, the FPD system \cite{lit:UWvalves} takes them from the exit of the MS to the primary KATRIN detector, a 148-pixel p-i-n-diode array on a monolithic silicon wafer. The dartboard-pattern pixelation scheme allows the separate analysis of different regions of the analyzing plane. The system contains electron and gamma calibration sources, as well as two superconducting solenoids to complete the MAC-E filter of the MS and to focus electrons onto the detector wafer. A post-acceleration electrode allows the signal electron energy to be elevated by up to 10 keV. The FPD vacuum system is divided into two independent regions: an external high-vacuum region containing the front-end electronics, and an internal UHV region, which contains the detector and couples to the MS vacuum via an all-metal DN250 gate valve followed by an in-beam valve (see Sec.~\ref{sec:vacuum_valve}). After roughing and bakeout, the vacuum in each region is maintained by a dedicated cryopump.


\section{The vacuum system of the Main Spectrometer \label{sec:vacuum}}

\begin{figure} 
\centering
\includegraphics[width = \textwidth]{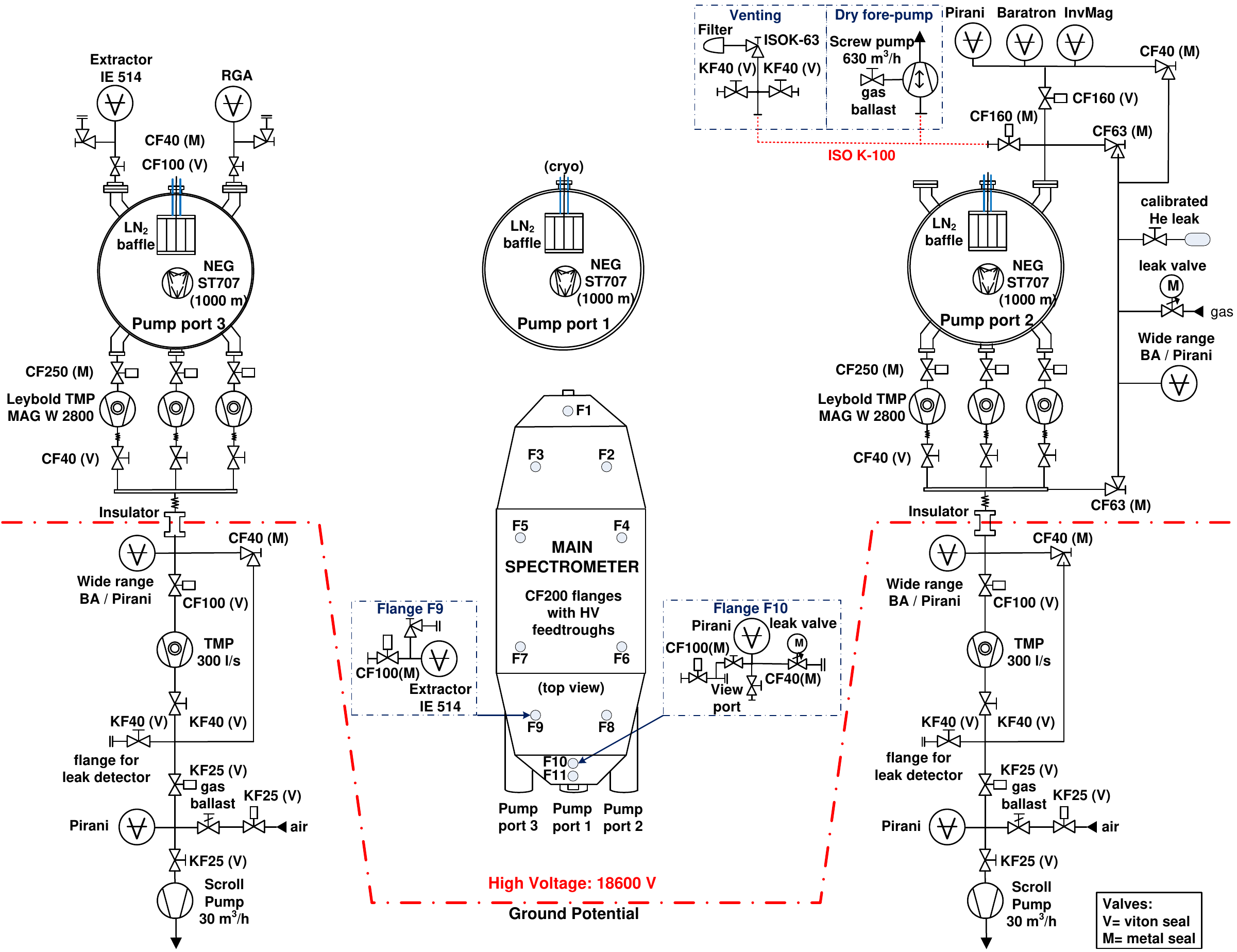} 
\caption[Vacuum scheme of the Main Spectrometer]
{Vacuum scheme of the Main Spectrometer. The Extractor gauges on
flange F9 (vessel) and on pump port P3 were used in the vacuum analysis.}
\label{fig:vacuum_scheme} 
\end{figure}

\subsection{The vacuum vessel\label{sec:vessel}}

The design goals of the Main Spectrometer vacuum system are to maintain a pressure
in the lower $10^{-11}\,$mbar regime during the entire 5-year lifetime of the KATRIN
experiment, as well as suppressing tritium and radon induced background. The
vessel has a volume of 1240$\,$m$^3$ and an inner surface area of 690$\,$m$^2$.
The inner wire electrode system, with a total of 120,000 individual parts,
adds another 532$\,$m$^2$ of stainless steel parts to the inner surface, increasing the total surface area 
to 1222$\,$m$^2$. An overview of the principal construction materials and the respective 
surface areas of the MS components is given in Table~\ref{tab:surfaces}. 

The outgassing rate of this large surface is the limiting factor for the ultimate pressure in the MS. 
Measurements with the pre-spectrometer showed that
a hydrogen outgassing rate of $10^{-12}\,\rm{mbar}\cdot \ell/\rm{s\cdot cm}^2$ can
be reached for 316LN stainless steel, after electro-polishing, cleaning, and 
vacuum baking at temperatures of at least 200\,$^\circ$C.
The cleaning process of the MS involved
pickling, electro-polishing and rinsing with an alkaline
degreaser and deionized water. Methods to reduce
hydrogen outgassing, such as vacuum firing at high temperatures,
could not be applied due to the size of the
MS \cite{lit:JVS09}. Most other components, installed inside the MS, were cleaned in similar processes, 
using ultrasonic baths for the alkaline degreaser and the deionized water. In a final step they 
were dried in a drying oven at 110$\,^\circ$C for about 12 hours.   

Three tubular pump ports, each with a diameter of 1.7$\,$m and
a length of approximately 3$\,$m (see Fig.~\ref{fig:mainspec}), 
protrude from the detector-facing end of the main vessel. Each pump port is closed with a 
DN1700 flange, developed and tested with the PS \cite{lit:PS_outgassing}. 
The flanges are metal-sealed by custom-made 
double-gaskets. If the innermost gasket developed a leak, the volume 
between the two gaskets can be pumped down, reducing the leak-rate by up to five orders of magnitude.
So far this backup solution has not been needed, since the inner gaskets always stayed leak-tight at temperatures 
ranging from -20\,$^\circ$C to 350\,$^\circ$C. The spring-energized inner gaskets are 
made of silver-coated stainless steel tubes (type 321, diameter 9\,mm), bent to a ring, with both ends 
welded together \cite{lit:HTMS}. Inside each tube a stainless steel (type 302) spiral spring 
reinforces the tube and provides enough elasticity to allow movements of up to 0.2\,mm 
between the flanges. The outer gaskets are only made of type 321 stainless steel tubes 
without an internal spiral spring, since the requirements on the leak-tightness are less stringent.
The end-cap flanges of the two outer pump ports feature six DN400 knees
each, ending with DN250 Conflat (CF) flanges for TMPs, vacuum gauges and
feedthroughs for LN$_2$ cryogenic lines.
The electron beam-lines at the ends of both spectrometers connect to DN500 flanges, which use the
same flange design with spring-loaded metal gaskets. 

On the upper half of the MS
vessel eleven 50-cm-long DN200 ports with CF flanges (see Fig.~\ref{fig:mainspec} (left)
and the vacuum scheme in Fig.~\ref{fig:vacuum_scheme}) provide access to the inner electrode system.
On top of the ports 25-cm-long six-way crosses are mounted. The top port of each cross is sealed with a 
DN200 blank flange, while the four lateral ports with DN40 flanges provide access for feedthroughs for 
high voltage, internal temperature sensors, and a vacuum gauge (port F9). One of the DN200 blank flanges  
has been replaced by a gate valve, a sapphire window for laser measurements, and a remotely 
controlled leak valve for background measurements at elevated pressure 
(see Fig.~\ref{fig:vacuum_scheme}, port F10). Another blank flange has been replaced by a  
burst disk rated to 500$\,$mbar.

\begin{table}
\caption{The components inside the Main Spectrometer.  This table lists the construction materials of these 
components, as well as their surface area and nominal operating temperature. The cold baffles and the 
NEG strips are not included in the calculation of the outgassing rate. \label{tab:surfaces}}
\begin{center}
\begin{tabular}{lrrr}
\hline
{\bf component} & {\bf material} & {\bf temperature} & {\bf surface}  \\
\hline
MS vacuum vessel        & 316\,LN         & $20\,^\circ$C & 690.0\,m$^2$ \\
wire electrodes             & 316\,L           & $20\,^\circ$C & 472\,m$^2$  \\
electrode rail system     & 316\,LN         & $20\,^\circ$C & 58\,m$^2$ \\
feedtrough flanges       & 316\,LN         & $20\,^\circ$C & 2\,m$^2$ \\
ceramic insulators        & Al$_2$O$_3$ & $20\,^\circ$C & 6\,m$^2$ \\
anti-penning electrode  & Ti & $20\,^\circ$C & 11\,m$^2$ \\
ground electrodes & Al & $20\,^\circ$C & 1\,m$^2$ \\ \hline
cryogenic baffles & Cu &  $-187\,^\circ$C & 31\,m$^2$ \\
NEG strips & St 707    & $20\,^\circ$C & 180\,m$^2$ \\
\hline
\end{tabular}
\end{center}
\end{table}

\subsection{The vacuum pumps\label{sec:pumps}}

\begin{figure} 
\centering
\includegraphics[width = \textwidth]{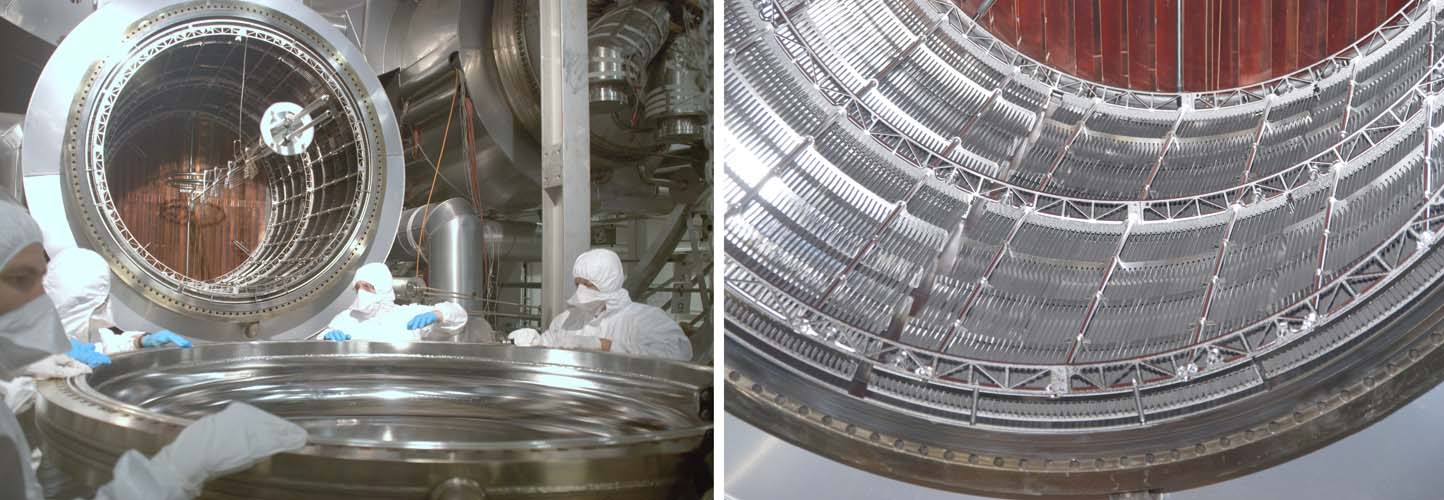} 
\caption{Left: Closing of one of the main pump
ports with the NEG pump and the cryogenic baffle visible in the background. Right:
A closeup view of the NEG pump with SAES St707 NEG strips installed. } \label{fig:pumpport}
\end{figure}

Figure~\ref{fig:vacuum_scheme} shows an overview of the vacuum system. 
Three custom-made NEG pumps \cite{lit:MS_NEG} (see Fig.~\ref{fig:pumpport}), 
each consisting of 1,000 1-m long SAES St707\textsuperscript{\textregistered} getter strips, 
are mounted on the three large pump ports.
The NEG strips are 30-mm-wide Constantan\textsuperscript{\textregistered} strips,
which are coated on both sides with a 27-mm-wide NEG area. 
Their combined nominal pumping speed for hydrogen is 1,000\,m$^3$/s
\cite{lit:SAES}. As mentioned before, the installation of LN$_2$-cooled baffles
in front of each getter pump is necessary for reducing the
radon-induced background component by capturing Rn atoms on the cold
surfaces. The conductance of the baffles (157\,m$^3$/s each) reduces the
total effective pumping speed of the three fully activated NEG pumps for hydrogen to 375\,m$^3$/s.
With an expected outgassing rate for stainless steel of
$10^{-12}\,$mbar$\cdot \ell$/s$\cdot$cm$^2$ \cite{lit:JVS09,lit:PS_outgassing}, the
ultimate pressure in the main volume would be $3.2\cdot 10^{-11}\,$mbar.

The two outer pump ports are each equipped with three TMPs (Leybold
MAG-W-2800\textsuperscript{\textregistered}), which use magnetic bearings in the rotor mechanism. 
They provide a combined effective pumping
speed for hydrogen of 10\,m$^3$/s. These pumps have three tasks:

\begin{itemize} 
\item initial pump-down during commissioning and baking of the
vacuum vessel, 
\item pumping of released hydrogen during NEG activation and 
\item pumping of non-getterable gases, such as noble gases and methane during 
standard operation. 
\end{itemize}

The fore-vacuum of each set of three TMPs is produced by a 300$\,\ell$/s TMP
backed by a scroll pump. This cascaded setup provides a high enough
hydrogen compression ratio for the MS to reach the required pressure regime.

As a roughing pump for the initial pump down a Leybold
SP630\textsuperscript{\textregistered} screw pump (630$\,$m$^3$/h) is
temporarily connected to one of the pump ports. After reaching $10^{-2}\,$mbar, 
the TMPs take over, and reduce the pressure to approximately $10^{-7}\,$mbar. 
This process takes two to three days. After reaching this pressure 
at room temperature the vessel has to be baked at
temperatures up to 350$\,^\circ$C to get rid of water and other contaminants on
the inner surfaces, and to activate the NEG pumps. After baking the system,
hydrogen outgassing from these surfaces is the limiting factor for the ultimate
pressure. A detailed description of the bake-out procedure is given in Section
\ref{sec:commissioning}.
 
\subsection{The vacuum gauges\label{sec:gauges}}

The intermediate vacua between the
cascaded TMPs and the fore-vacua, provided by the scroll pumps, are 
monitored by several wide-range gauges.
The ultra-high-vacuum (UHV) pressure level inside the vessel is measured by three
ionization gauges and one quadrupole mass spectrometer:

\begin{itemize} 
\item a Leybold Extractor gauge
(IE514\textsuperscript{\textregistered}) at the DN200 port F9 
(see Fig.~\ref{fig:vacuum_scheme}) on the
shallow cone at the detector end of the main volume, 
\item a Leybold Extractor gauge at a
DN400 port of pump port P3 behind the NEG pump, 
\item a MKS Inverted Magnetron
gauge (HPS 421\textsuperscript{\textregistered}) at a DN400 port of pump port P2,
serving mainly as a crosscheck during bake-out and at pressures above
$10^{-5}\,$mbar, and 
\item a MKS Microvision II\textsuperscript{\textregistered} quadrupole mass
spectrometer (RGA) at a DN400 port of pump port P3. 
\end{itemize}

Both Extractor gauges have been calibrated against a Bayard-Alpert reference gauge ($\pm 10\%$) for 
nitrogen, hydrogen, helium and argon
at $10^{-6}\,$mbar. An in situ calibration at low pressure in the Main
Spectrometer was not performed due to the scheduling of the electromagnetic test measurements 
with the MS. The residual gas in the spectrometer was a mixture of 
different gases. Therefore the pressures are reported, using the  
nitrogen calibration, if not mentioned otherwise. Gas correction factors relative to the nitrogen 
calibration have to be applied, if one gas type dominates the 
mixture. The factors have been determined for both Extractor gauges (F9, P3)
for hydrogen (2.2, 2.3), helium (5.6, 5.7), and argon (0.70, 0.71).  
Another correction factor of 1.06 had to be applied to the Extractor gauge at pump port 3
for the time intervals when the superconducting magnets were turned on, since the sensitivity of ionization gauges 
is influenced by magnetic fields. The factor was measured by comparing the displayed
pressure before and after the magnetic field was switched on.

While the gauge at P3 was connected to the pump port volume through a 400-mm-diameter tube, the 
gauge at F9 was only connected to the main volume through a 40-mm-diameter and 25-cm-long tube, 
ending in the 50-cm-long DN200 port. With its filaments 
close to the walls of the tube, and a much lower conductance compared to P3, 
the lowest pressure measured in F9 was limited by local outgassing, which produced a measured offset in 
the order of $10^{-10}\,$mbar.   

Before and during the bake-out the RGA was used with the built-in Faraday cup detector.
At low pressure the sensitivity of the RGA was increased by using the built-in 
secondary electron multiplier (SEM) detector.
The RGA peaks were normalized against the nitrogen-calibrated signal of 
Extractor gauge P3, which was mounted at a similar location as the RGA in pump port 3. 
For two gas species an additional correction factor was applied to the peaks: 
hydrogen (mass 1, 2 and 3), and argon (mass 40, 36, and 20). 
For all other mass peaks the nitrogen calibration of the Extractor gauge was used. For more details 
see Appendix~\ref{sec:RGAcalibration}. The RGA signal was very sensitive to magnetic fields.
Therefore both the Extractor gauge and the RGA were passively shielded by two layers of soft iron tubes that are 
fabricated from Fe-360 metal sheets. Even with this shielding the magnetic fields affect the 
amplitude of the measured mass peaks by 10\,\% to 20\,\%, depending on the mass.  
The effects for lower mass peaks are more pronounced. 
Therefore the RGA spectra, shown in this paper, were taken at times 
when the magnetic fields were off.

One of the Extractor gauges is mounted at the main volume of the vessel (port F9), while the other one 
is located in pump port P3 behind the baffle and the NEG pump. Therefore, they measure different pressure
values. Their pressure ratio depends on the pumping speed of the NEG pump and on the gas
composition. Based on vacuum simulations, described in the following sections, 
the pressure ratio between these two
gauges can be used to estimate the level of activation of the getter pumps (see
Section \ref{sec:NEGpump}). During standard operation with the high voltage
switched on, the Extractor gauge at the main vessel (port F9) is switched off,
since it produces background electrons that would interfere with the low count-rate 
measurements. In this case, the pressure inside the main vessel can only be
estimated from the value of the Extractor gauge at pump port 3, which is
usually a factor of 2 to 5 lower than the actual pressure in the main volume, 
depending on the level of activation of the NEG pumps.


\subsection{The in-beam valves\label{sec:vacuum_valve}}

During commissioning and bake-out of the MS it is necessary to attach and remove hardware from either end of the spectrometer. To avoid exposing the conditioned spectrometer to atmosphere, custom valves were designed to satisfy several requirements. First, they must be able to temporarily seal the spectrometer at a maximum leak rate of $< 10^{-7}\,$mbar$\cdot\ell$/s helium through the valve. Since they must be attached to the MS during bake-out, they must tolerate temperatures of up to 200$\,^\circ$C, and they must accommodate up to 2\,cm thermal movement of the spectrometer in addition to spatial adjustments for beam alignment. Finally, when in the open position, the valves must provide unobstructed passage for the electron beam.

Due to the movements of up to 12\,cm of the MS vessel, caused by thermal expansion 
and contraction during the bake-out, 
both the electron gun and the detector system are disconnected before the start 
of the bake-out, with both in-beam valves closed. The valves are not intended to provide a long-term solution to isolating the MS. Once the valve is closed, the volume exposed to atmosphere is capped with a CF blanking flange and then evacuated.

\begin{figure} 
\centering
\includegraphics[width = 0.7\textwidth]{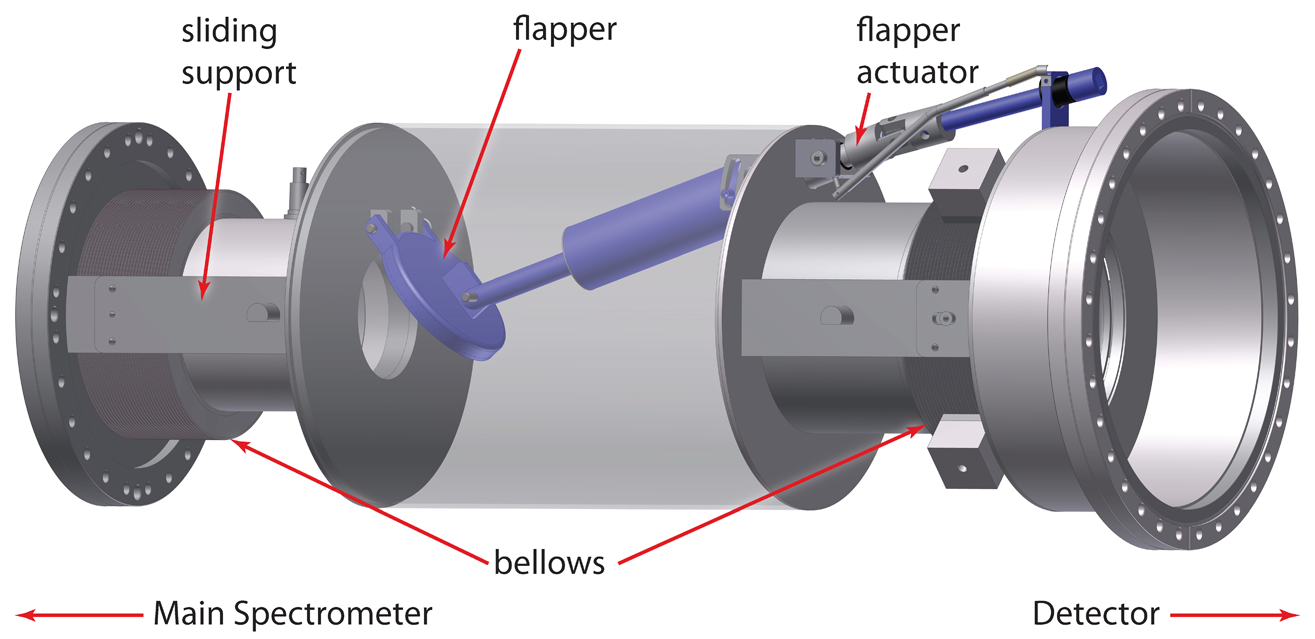} 
\caption{View of the engineering model for the in-beam valve separating the detector system from the MS. The central body of the valve is transparent to show the flapper mechanism. } \label{fig:beamlinevalve}
\end{figure}

An engineering model of the valve separating the MS from the detector system is shown in Fig.~\ref{fig:beamlinevalve}. The valve between the PS and the MS is similar. Space constraints demanded that the valves fit inside the warm bores of the superconducting magnets at either end of the spectrometer. Edge-welded metal bellows at either end of the valve accommodate MS movement and facilitate alignment. Sliding joints support the body of the valve without impeding the movement of the bellows. The valve closure is a simple, manually operated flapper mechanism. The flapper is sealed by a Kalrez\textsuperscript{\textregistered} O-ring which was chosen for its lower radon emanation compared to Viton\textsuperscript{\textregistered} O-rings \cite{lit:RadonHS}.
 
\subsection{Vacuum and spectrometer operation\label{sec:vacuum_SDS}}

During pump down and baking the MS is at ground potential, while for standard
operation it is at high potential. Therefore the whole vessel is supported on
electrical insulators, on which it can slide freely during the bake-out. All vacuum
devices and other equipment directly mounted on the vessel have to be connected
to controllers, which are installed in electrically insulated cabinets and powered
via an isolation transformer. At both ends of the MS the beam-line is connected
through 171-mm-long conical Al$_2$O$_3$ ceramic insulators, which are mounted at the
central DN500 flanges. Each set of three TMPs at the pump ports P2 and P3 is isolated 
from the grounded fore-vacuum system by a 200-mm-long DN100 ceramic tube 
on the high vacuum flange of the 300-$\ell$/s TMP.  In order to prevent
gas discharge, the control system switches off the high voltage if the pressure
in one of the insulators rises above $10^{-4}\,$mbar.

The static stray magnetic fields of the superconducting solenoids and air-coils cannot only influence the
vacuum gauges, but also the TMPs. The fast-moving all-metal
rotors of TMPs are susceptible to heating by eddy currents, induced by 
the external magnetic fields. 
The maximum field strength at the location of the MAG W 2800 TMPs is 1.7$\,$mT.
The expected rotor temperature of 65$^\circ$C is acceptable and can be tolerated
without countermeasures, as extensive tests have shown \cite{lit:TMP1, lit:TMP2}.

For the first electro-magnetic test measurements between May and September 2013
a high precision, angular selective electron gun \cite{lit:egunMS} was
used, which sent electrons through the spectrometer for counting by the detector at the other end. 
After bake-out the electron gun and the detector system were connected to the in-beam valves, 
and evacuated down to approximately $10^{-10}\,$mbar, before opening the valves to the MS.


\section{Simulation of the vacuum system \label{sec:simulation}}

The performance of the vacuum system has been estimated by
detailed simulations of the spectrometer vessel using Molflow+ 2.4 \cite{lit:molflow, lit:kersevan}.
Molflow+ is a Test Particle Monte Carlo (TPMC) code for simulating vacuum systems in the molecular flow regime. 
Three different parameters have been extracted from the simulations for hydrogen and radon: 
(i) the effective pumping speeds for the TMPs, NEG pumps, and baffles, 
(ii) the conductance of a baffle, and  
(iii) the pressure ratio at the locations of the Extractor gauges.

After a particle is started in Molflow+ (desorbed from a surface), it is tracked through the geometry, 
until it is adsorbed on a pumping surface with a certain sticking coefficient $\alpha$, 
defined as the probability that a particle sticks to the surface after 
impinging on it. All other particles are diffusely reflected.
Pumps are simulated by one or more surface elements with
appropriate values for $\alpha$. The pumping speed $S$ of a pump with an opening area $A$ 
of the high vacuum flange, for 
gas particles with an average speed $\bar{c}$ (H$_2$: 1754.6$\,$m/s; Rn: 167.2$\,$m/s), is

\begin{equation}
S = \frac{1}{4}\bar{c}\cdot A \cdot w. \label{equ:pumping-speed}
\end{equation}

The parameter $w$ is the pumping probability, defined as the ratio of particles absorbed by the pump, 
to the particles entering the pump through the opening $A$ (for instance the high vacuum flange of a TMP). 
This definition is similar to $\alpha$, but for a more complex geometry. Particles leaving the pump towards 
the vacuum vessel are discarded by setting the sticking coefficient of the entrance $A$ to 100\%. 

The conductance of a component  with two openings (for instance a tube) is the product of the 
transmission probability $w$ and the flow $\dot{V} = 0.25\cdot \bar{c}\cdot A$ into 
the component through opening $A$. The transmission probability is defined as the ratio of particles
entering through opening $A$ and leaving the component through the other opening. 
The sticking coefficients are set to 
100\% for both opening surfaces. Thus the simulation of a conductance is similar to the simulation 
of a pumping speed (Equ.~\ref{equ:pumping-speed}). 

The pressure at the location of a vacuum gauge is proportional to the number of particles hitting
a surface element of the model, divided by the area of the element. The absolute pressure cannot be 
simulated directly, but has to be calculated, using further information, such as the total desorption rate or 
the average speed $\bar{c}$. However, the ratio of two pressure values at different locations $i$ can 
be easily calculated from the number of hits $H_i$ and the respective areas $A_i$:
\begin{equation}
\frac{p_1}{p_2} = \frac{H_1}{H_2}\cdot\frac{A_2}{A_1}. \label{equ:pressure-ratio}
\end{equation}

Two models were set up to describe the Main Spectrometer. 
Model 1 comprised the whole spectrometer vessel, including baffles, 
NEG pumps, TMPs, and vacuum gauges. Particles were started from all surfaces, assuming 
homogeneous outgassing rates and a cosine angular distribution. This model was used to simulate 
pressure ratios of the two Extractor gauges. 
Model 2 simulated one pump port (P3) with a baffle, a NEG pump, and three TMPs. All particles were 
started towards the pump port from the virtual surface of the intersection between the pump port and 
the main vessel. The sticking coefficient of this surface was set to 100\%, i.e. adsorbing all particles that
left the pump port towards the main vessel. This model was used to simulate the effective pumping
speeds of TMPs, NEG pumps, and baffles, respectively.

\begin{figure}
\centering
\includegraphics[width=\textwidth]{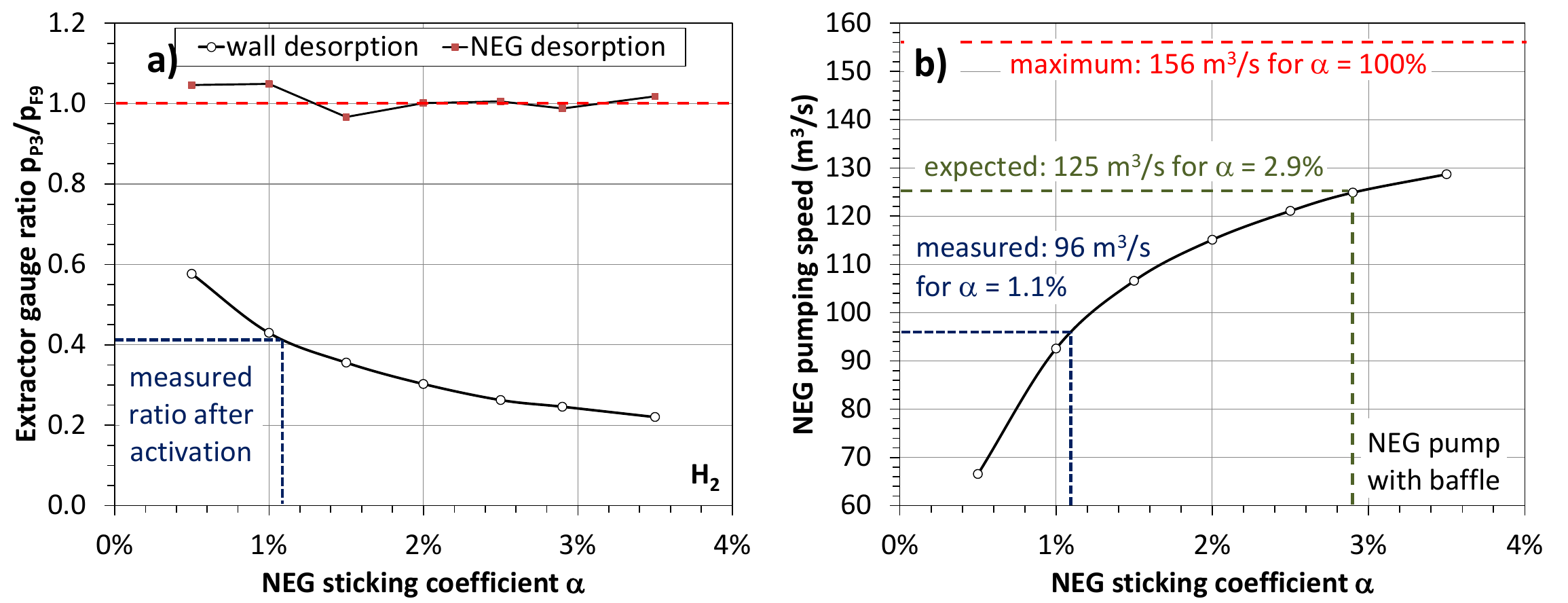} 
\caption[NEG simulation]{Molflow+ simulations of the Main Spectrometer with a NEG pump and a baffle
in each pump port. (a) Pressure ratio in pump port P3 behind the
NEG pump, and in the main volume of the spectrometer as a function of the NEG
sticking coefficient for hydrogen. The measured, hydrogen dominated ratio of
0.41 leads to a sticking coefficient of $\alpha = 1.1\%$. For a fully activated
NEG pump $\alpha = 2.9\%$ is expected. (b) For the partly activated NEG pump the
effective pumping speed for hydrogen was 96$\,$m$^3$/s in one pump port.
}\label{fig:simulation_NEG} 
\end{figure}

\subsection{Simulation of hydrogen \label{sec:simulation_H2}}

The NEG pumps were defined as
1000$\,$m of getter strips in each pump port, with a sticking coefficient $\alpha$, varying from  0.5\%
to 3.5\% in 7 simulations. Since only 27\,mm of the 30-mm-wide real getter strips are coated with NEG material, 
the sticking coefficients in the simulated strips were reduced by 10\%.  
The simulated pressure ratio (model 1) between the Extractor gauges
in the pump port behind baffle and NEG pump (P3), and in the main volume (F9) is shown in
Fig.~\ref{fig:simulation_NEG}.a. This plot is used in Section~\ref{sec:NEGpump} 
to determine the actual sticking coefficient of the NEG strips after activation. If the NEG pumps are 
not activated 
($\alpha = 0$), gas is only pumped by the TMPs. In this case the effective pumping speed is small
compared to the conductance of the baffles, and the pressure ratio converges towards 
$p_{\rm P3}/p_{\rm F9} = 1$.

In a second simulation with the same model, desorption was only defined for the surfaces of the NEG strips. 
This corresponds to the situation during activation of the NEG pumps, when hydrogen is driven out 
of the getter and dominates the pressure distribution in the Main Spectrometer. Since gas
can only be pumped through the pump ports, the gas flow into the main volume continues until an equilibrium
with the pump ports is reached. The pressure ratio between the Extractor gauges is approximately $p_{\rm P3}/p_{\rm F9} = 1$.  

The effective pumping speed of the NEG pump in pump port 3 was simulated using model 2.
The pumping probability $w$ of the NEG pump was
calculated as the number of particles adsorbed on the surfaces divided by the total number 
of particles desorbed from the virtual surface of the intersection between pump port 3 and the main vessel. 
The effective pumping speed of the NEG pump was determined by varying the sticking probability 
$\alpha$ between  0.5\% and   3.5\% in seven steps (see Fig.~\ref{fig:simulation_NEG}b).  
With the conductance of the baffle (simulation: $157\,$m$^3$/s) as the limiting factor, the effective 
pumping speed for the fully activated getter strips in a pump port ($\alpha = 2.9\%$ \cite{lit:SAES}) 
was $125\,\rm{m}^3$/s, or $375\,\rm{m}^3$/s for all three pump ports of the Main Spectrometer.

The simulation of the effective pumping speed without the baffles (as initially planned) resulted in a
value of $930\,\rm{m}^3$/s for the spectrometer. Thus, the necessity of introducing the baffles for
capturing radon atoms reduced the effective pumping speed for hydrogen by 60\%.

\subsection{Simulation of radon \label{sec:simulation_Rn}}

Radon was simulated for two different sources \cite{lit:RadonFF}: 
(i) NEG strips are known to emanate small amounts of $^{219}$Rn with a half-life of 4.0$\,$s, and 
(ii) the stainless steel walls and weldings of the vessel emanate small amounts of  $^{219}$Rn and 
$^{220}$Rn with a half-life of 55.6$\,$s. The most common radon isotope,  
$^{222}$Rn with a half-life of 3.8$\,$d,
which might also be emanated, is not taken into account, since almost all of it is pumped out 
before it decays.

\begin{figure}
\centering
\includegraphics[width=\textwidth]{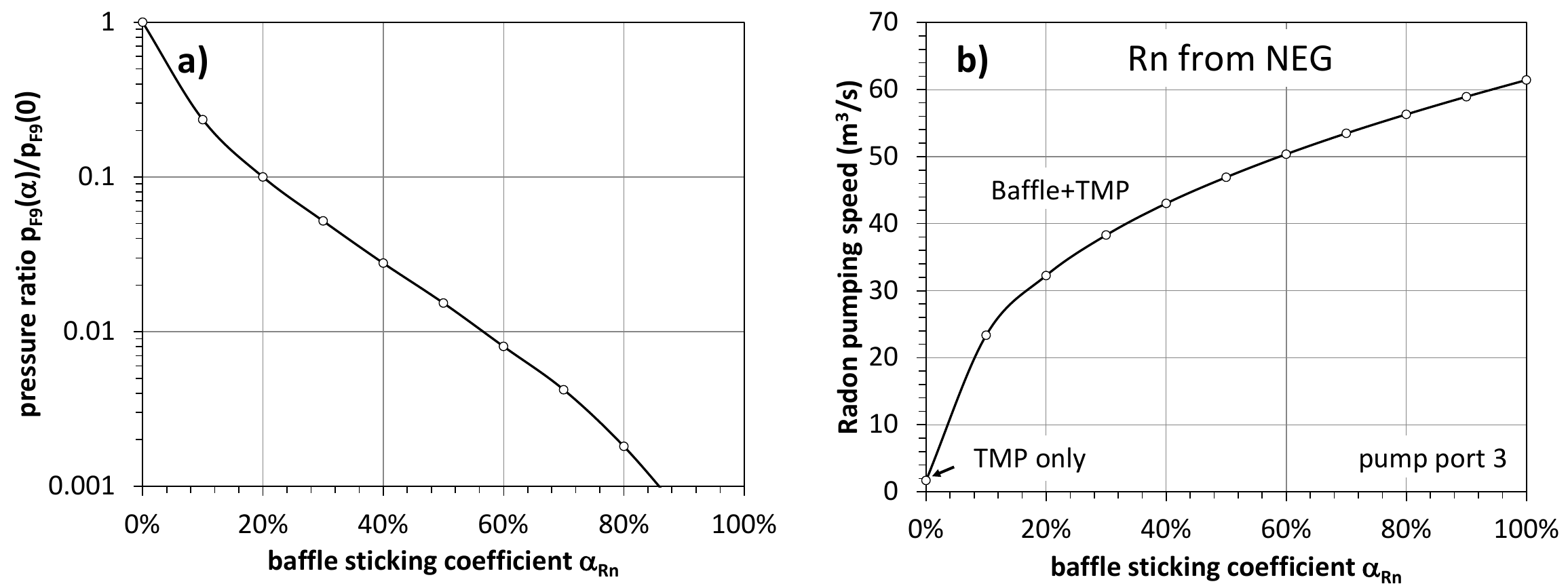} 
\caption[Radon baffle simulation]{Simulation of the 
suppression factor of the cryogenic baffles for
radon emanating from the NEG strips in the pump ports (a), and effective pumping
speed of the baffles for radon emanating from the inner surfaces of the
spectrometer (b), plotted over the sticking coefficient for radon. A sticking
coefficient of 80\% is estimated for nitrogen-cold baffles, using previous
measurements with the pre-spectrometer. }\label{fig:Rn_simulation} 
\end{figure}

Radon, emanated from the NEG strips in the pump ports, has to be prevented from 
entering the spectrometer by the cryogenic baffles
\cite{lit:phd:mertens,lit:phd:wandkowsky}. The suppression factor of the baffles 
for radon has been simulated using model 1. 
The pressure ratio $p_{\rm P3}/p_{\rm F9}$ of the two Extractor 
gauges serves as the measure for radon suppression. This number takes into account that a 
considerable fraction of the radon atoms already decay inside the spectrometer volume. The sticking 
coefficient for radon, which strongly 
depends on the temperature of the baffles, was varied from 0\% to 90\% in the calculations. 
The results are shown in
Fig.~\ref{fig:Rn_simulation}a. From the measurements with the pre-spectrometer we expect the 
sticking coefficient to be $\sim$80\%, which would result in a radon suppression factor of 550.

Radon emanated directly into the main volume from its inner surfaces cannot be prevented from decaying 
in the flux tube. It can only be pumped out quickly enough before a large fraction can decay
 \cite{lit:RadonSM}.
The pumping speed of a baffle in pump port 3 with regard to the main volume has been simulated, using model 2.
 Fig.~\ref{fig:Rn_simulation}b 
shows the results for the simulated sticking coefficients ranging from 0\% to 100\%. For a sticking coefficient 
of 80\% the effective pumping speed is  56$\,$m$^3$/s, resulting in a total pumping speed of 
$S = 170\,$m$^3$/s.
The pump-out rate $S/V = 0.14\,s^{-1}$ has to compete with the decay rates for $^{219}$Rn
($\lambda_{219}=0.17\,s^{-1}$), and $^{220}$Rn ($\lambda_{220}=0.012\,s^{-1}$), as well as
re-desorption from the baffles \cite{lit:phd:goerhardt}.


\section{Commissioning of the vacuum system and status after bake-out
\label{sec:commissioning}}

\subsection{Pump down and leak tests  \label{sec:pump_leak}}

Commissioning of the Main Spectrometer vacuum system started in summer 2012,
after a four-year period when the complex inner electrode system, 
the cryogenic baffles, and the getter pumps were installed under cleanroom
conditions. After the initial pump-down with the SP630 screw pump and three TMPs on
pump port 3, the vessel vacuum reached $10^{-7}\,$mbar. Pump down of the vessel 
to $10^{-2}$\,mbar took 6 days due to some coarse leaks, which had to be closed first, 
before the TMPs could be switched on. A final leak test with a sensitivity of $5\cdot
10^{-10}\,$mbar$\cdot \ell$/s was performed with a leak detector used as fore-pump
for the three TMPs. With an effective pumping speed for helium of 6000$\,\ell/s$
and a total volume of 1240$\,$m$^3$ the response time\footnote{Time to reach
90\% of the signal.} of each leak test was 7$\,$min. For each local leak test
the respective flange was enclosed in a tightly sealed plastic bag filled with helium.

\subsection{Bake-out procedure\label{sec:baking}}

The nominal operating temperature of the spectrometer vessel is $20\,^\circ$C.
If needed, it can be cooled down to $10\,^\circ$C, thus reducing the H$_2$
outgassing rate of stainless steel by a factor of two. During vacuum bake-out it
can reach temperatures up to $350\,^\circ$C. The temperature is controlled to
better than $1\,^\circ$C by a thermal oil temperature unit from
HTT\textsuperscript{\textregistered} \cite{lit:HTT}. The system has two
independent thermal circuits, one for the main vessel and one for the three pump
ports with the getter pumps. It provides a total heating power of 440$\,$kW and
a cooling power of 60$\,$kW using 9$\,$m$^3$ of heat transfer fluid (Marlotherm
LH\textsuperscript{\textregistered} \cite{lit:marlotherm}). The fluid is
continuously pumped through approximately $1200\,$m of 114-mm-diameter half-tubes 
welded to the outer surface of the spectrometer vessel. In
addition the heating system is supported by 56 electrical heating tapes for
smaller ports, flanges and gate valves. The temperature distributions on the exterior 
surface of the MS, on the interior surface at the inner electrode system, at the getter pumps and at the
LN$_2$-baffles are monitored by an array of 381 temperature sensors. 
On the outer surface PT100 sensors are used. Inside the vacuum vessel temperatures are
monitored by PT1000 sensors, which are attached to the NEG pumps, the LN$_2$-baffles and 
some of the inner electrode frames.

\begin{figure} 
\centering
\includegraphics[width=\textwidth]{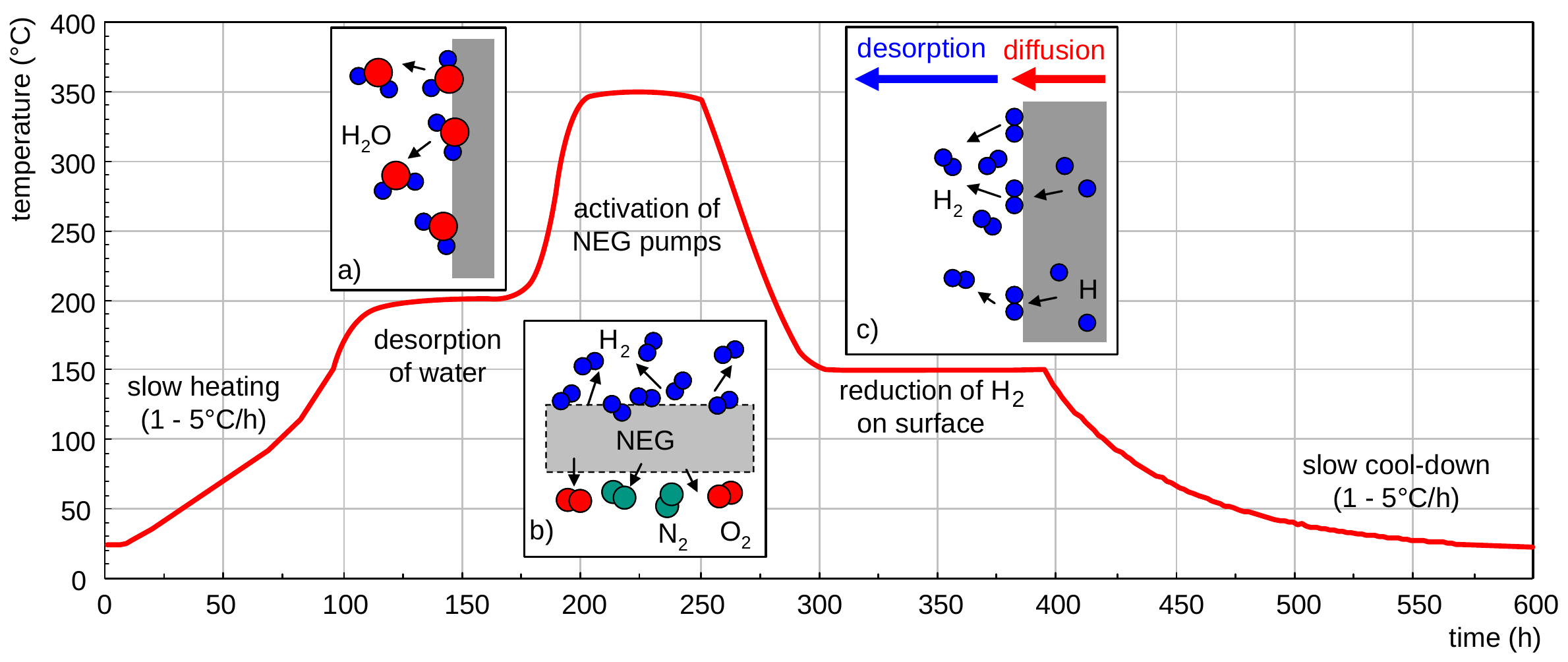} 
\caption[Main Spectrometer bake-out schedule]
{Based on test measurements with a $300\,\ell$
vacuum vessel, an optimized schedule for the MS bake-out cycle
was defined: \textbf{a)} accelerated desorption of water from all surfaces,
\textbf{b)} activation of the NEG pumps at $350\,^\circ$C, and \textbf{c)} intermediate
temperature step, in order to further reduce hydrogen on all surfaces. For all
temperature changes the gradient had to be kept in the range of 1 to
$5\,^\circ$C/h in order to allow the inner electrode system to follow the
temperature of the vessel within $2\,^\circ$C. } \label{fig:schedule} 
\end{figure}

The bake-out procedure was tested and optimized with a smaller vacuum vessel
(volume: $300\,\ell$), built with the same type of stainless steel (316LN) as
the MS. Based on these measurements a schedule for the MS
bake-out cycle was devised \cite{lit:phd:goerhardt} for both cleaning the inner 
surfaces and activating the getter strips at $350\,^\circ$C (see Fig.~\ref{fig:schedule}):

\begin{enumerate} 
\item Increase the temperature slowly to $200\,^\circ$C. Up to
a temperature of $90\,^\circ\mathrm{C}$ a ramping speed of $1\,^\circ$C/h is
used. Above this temperature the ramping speed is slowly increased in several
steps, with a maximum rate of $5\,^\circ$C/h above $150\,^\circ\mathrm{C}$.
\item Keep the temperature stable at $200\,^\circ$C for about two days, in order
to remove most of the water bound on the stainless steel surfaces and to reduce
the outgassing of hydrogen before activating the NEG pumps. 
\item Increase the
temperature to $350\,^\circ$C at a rate of $5\,^\circ$C/h. 
\item Keep the temperature
stable for at least 24 hours to activate the NEG pumps, as recommended by the
manufacturer. 
\item Lower the temperature to $150\,^\circ$C at a rate of $5\,^\circ$C/h. 
\item Keep the
temperature stable for at least one day or until there is no further significant
pressure drop over time. This step is expected to reduce the hydrogen
concentration on the surface by desorption, but without replenishing the hydrogen by
diffusion from the bulk. At this temperature tests achieved the lowest outgassing 
rate of $3.5\cdot10^{-13}\,\rm{mbar}\cdot\ell/\rm{s\cdot cm}^2$.  
Other temperature steps between $100\,^\circ$C and $350\,^\circ$C reached
outgassing rates from 5.1 to $11.4\cdot10^{-13}\,\rm{mbar}\cdot \ell/\rm{s\cdot cm}^2$. 
\item Lower the temperature to $20\,^\circ$C. At this
time the residual gas composition in the clean vessel is expected to be
dominated by hydrogen diffusion from the bulk, with small traces of water, CO and CO$_2$. 
\end{enumerate} 

The slow ramping
speed of the temperature is necessary to protect the complex inner electrode
system, allowing it to follow the temperature profile of the main vessel.
Between room temperature and $350\,^\circ$C the circumference of the Main
Spectrometer vessel expands by $15\,$cm, while the electrode modules can only
compensate movements of a few mm against the vessel. Since the electrodes are
mainly heated by radiation, the temperature gradient between vessel and
electrodes is constantly monitored. The slower rise time at low temperatures
takes into account the T$^4$ dependence of heat radiation. During the whole
bake-out and cool-down procedure the temperature gradient between vessel and
electrodes was kept at $1\,^\circ$C, ensuring a similar thermal expansion rate
of both systems.

\begin{figure} 
\centering
\includegraphics[width=0.9\textwidth]{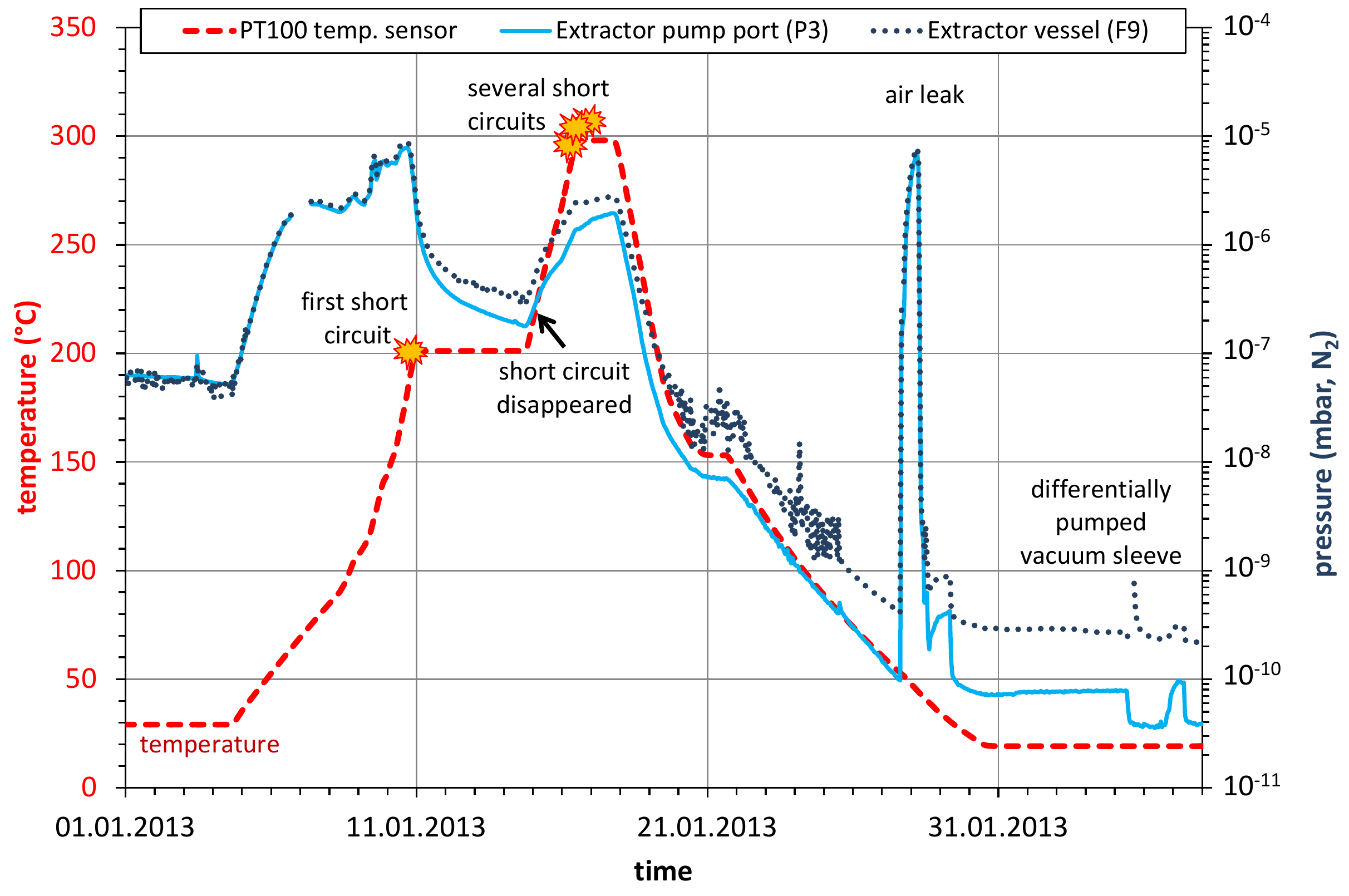} 
\caption[Main Spectrometer bake-out cycle - January 2013]
{Overview of the actual Main
Spectrometer bake-out cycle in January 2013. Plotted are pressure and
temperature versus time. The occurrences of short circuits in the electrode
system and an air leak during the final cool down phase at about $52\,^\circ$C
are indicated in the plot.} \label{fig:bake-out2013} 
\end{figure}

An overview of the whole bake-out cycle, which started on January 4$^{\rm{th}}$
and ended on January 31$^{\rm{st}}$, 2013, is given in
Fig.~\ref{fig:bake-out2013}. Within a time interval of 36$\,$h, when the
temperature rose from $120\,^\circ$C to $200\,^\circ$C, the water content in the
residual gas spectrum dropped by a factor of 50. During the following four days
the water content dropped by another factor of 5, while the vessel was kept at a
constant temperature of $200\,^\circ$C.

During the whole bake-out cycle different sections and wire layers of the
inner electrode system were constantly monitored for short circuits. At a
temperature of $200\,^\circ$C the first incident occurred. A short circuit
between two wire layers in one of the central sections was detected. After
ramping of the temperature started again, the short circuit initially
disappeared. However, between $250\,^\circ\mathrm{C}$ and
$300\,^\circ\mathrm{C}$ several more sections developed short circuits 
between wire layers and between adjacent sections. As later inspections through the CF
ports for the high voltage feedthroughs showed, these short circuits were caused
by the deformation of the CuBe rods that connect the distribution panels to the corners of some of
the modules underneath the feedthroughs. 
In order to prevent further damage, the bake-out procedure was
stopped at $300\,^\circ\mathrm{C}$. This temperature was sustained for 28 hours,
activating the NEG pumps at least partly before being reduced to
$150\,^\circ\mathrm{C}$. The deformation can be traced back to the 
fact that the rods lost their tensile strength at temperatures above 200$\,^\circ$C, started 
to move downwards, and remained in this deformed position.

\begin{figure} 
\centering
\includegraphics[width=0.5\textwidth]{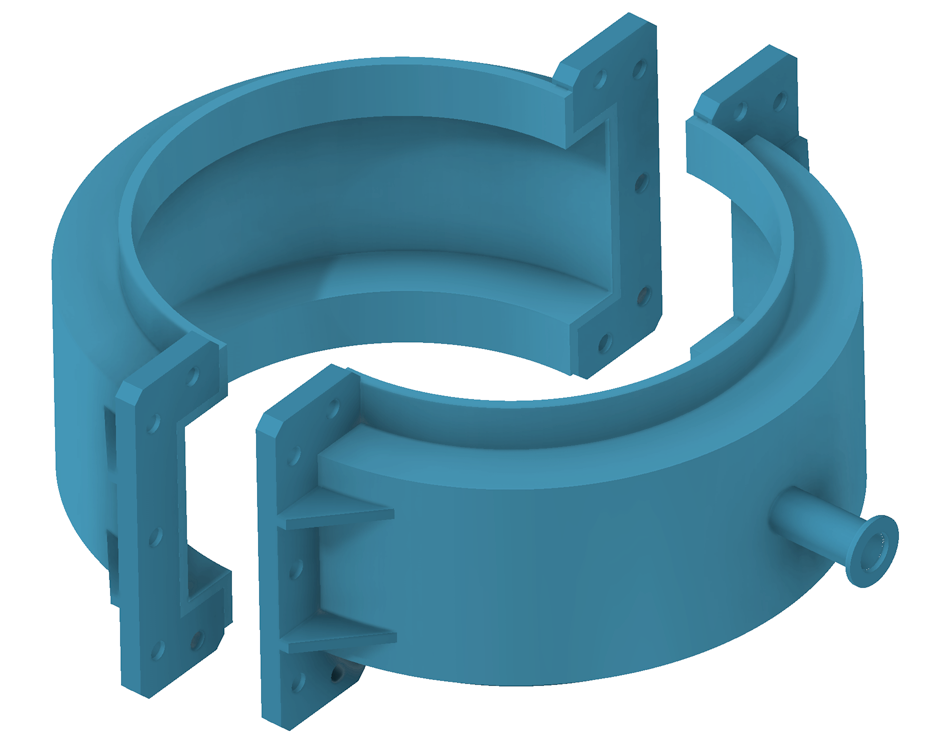} 
\caption[Air leak: vacuum sleeve]
{Two air leaks  opened up at CF flanges during the cool-down
phase after baking (port F8), and during pump-down after venting the vessel
with argon (port F9). Both leaks could be sufficiently reduced by differential
pumping (scroll pumps) using two pairs of vacuum sleeves, shown in this drawing, that were 
mounted around the leaking flange connections. }\label{fig:vacuum_sleeve} 
\end{figure}

In the final step the temperature was slowly reduced to $20\,^\circ\mathrm{C}$.
During this final cool-down a major air leak opened up at a temperature of
$52\,^\circ\mathrm{C}$ at a DN200 CF flange of port F8, which is on one of the conical
sections of the spectrometer. The pressure rapidly increased by five orders of
magnitude. Tightening the bolts of the flange was not sufficient to close the
leak, but an immediate repair of the leaking gasket would have resulted in a two-month 
delay and considerable operating costs for an additional baking cycle
after venting the spectrometer. Therefore a differentially pumped vacuum sleeve
(see Fig.~\ref{fig:vacuum_sleeve}) was installed around the flange and pumped
down to approximately 0.1$\,$mbar by a scroll-pump. This temporary measure
reduced the leak rate sufficiently to allow the continuation of the planned electro-magnetic test
measurements until October 2013, when the spectrometer was scheduled to be
vented again. A second leak opened up after the venting with argon (described in 
Sec.~\ref{sec:venting}). A similar vacuum sleeve was used to reduce this leak.

\begin{figure}
\centering
\includegraphics[width=0.9\textwidth]{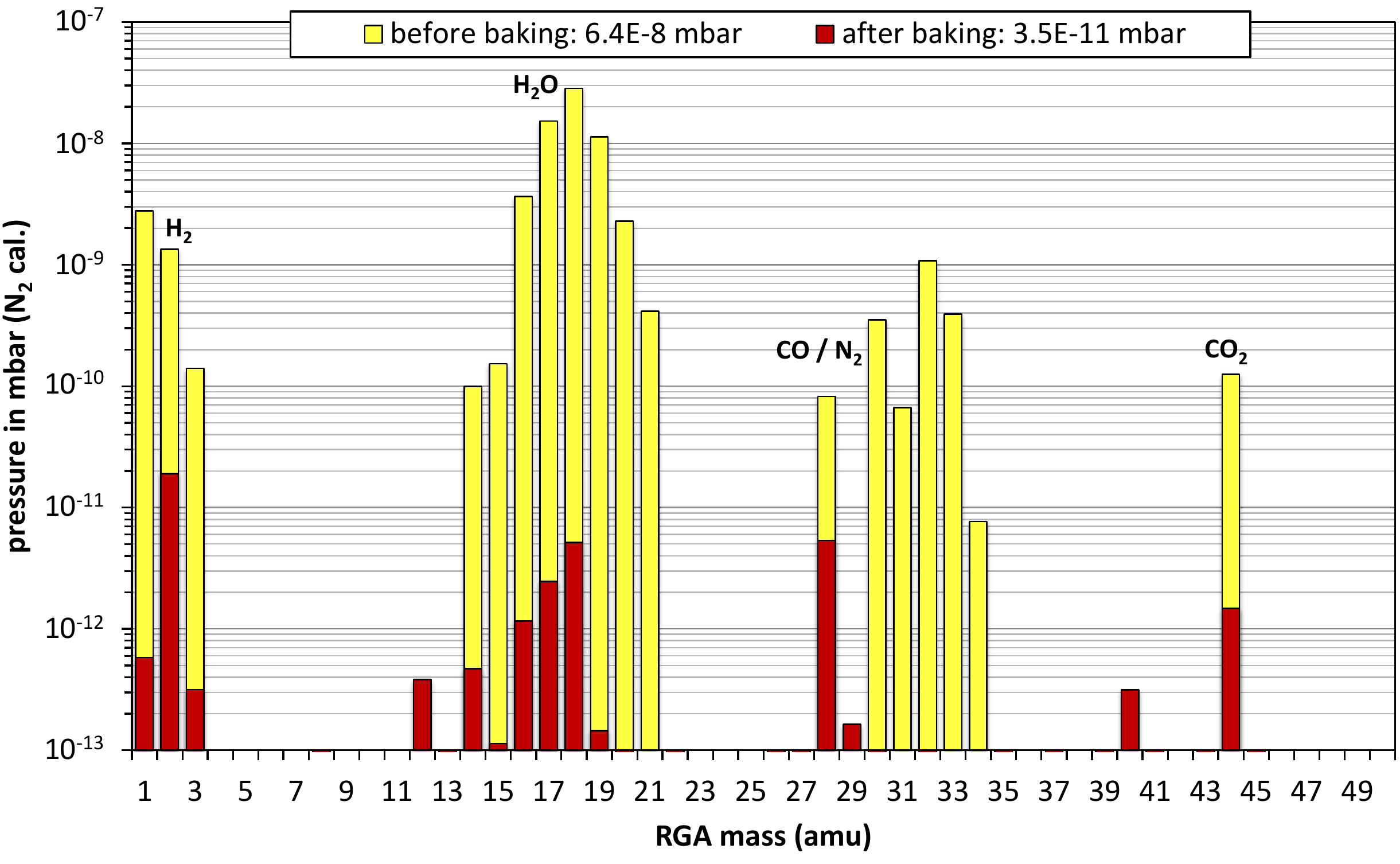}
\caption[Residual gas analysis - before and after bake-out]{RGA spectra before
and after bake-out. Before the bake-out procedure the pressure was dominated by
water (mass = 18$\,$amu). After the campaign hydrogen was the dominant residual
gas. The RGA peaks have been calibrated against the absolute pressure, as measured
with the Extractor gauge of pump port 3. The hydrogen pressure in the main
volume was approximately $6\cdot 10^{-11}\,$mbar (see text).}\label{fig:rga_baking} 
\end{figure}

\subsection{Vacuum performance after baking \label{sec:performance_bake}} %

After reaching the base temperature of $20\,^\circ\mathrm{C}$, the RGA spectrum (see
Fig.~\ref{fig:rga_baking}) revealed a hydrogen-dominated composition that also shows
traces of water, CO/N$_2$, and CO$_2$, as expected for a very clean vessel.
Despite the problems described in the previous section, the baking cycle reduced the 
final pressure by three orders of magnitude. 

\begin{figure}
\centering
\includegraphics[width=\textwidth]{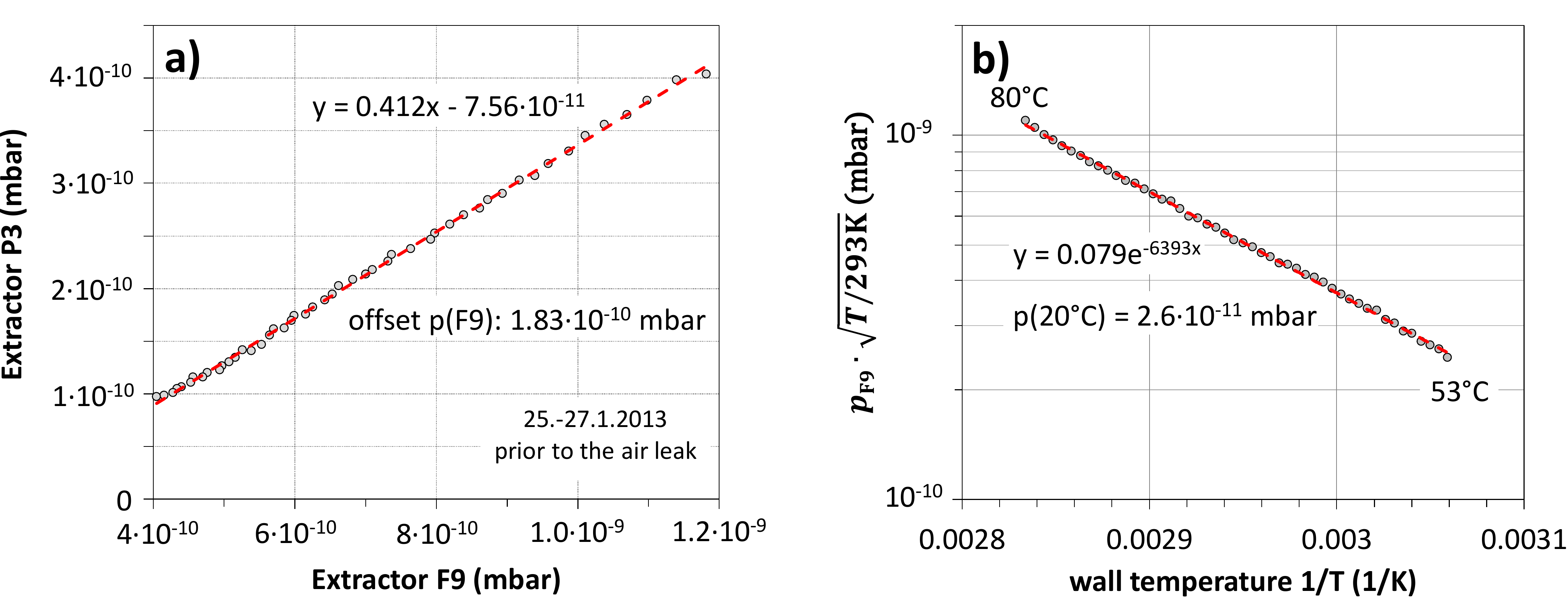}
\caption[Pressure ratio and Sievert plot]{Extractor gauges on port F9 (main volume) 
and pump port 3: a) Hydrogen dominated pressure, measured just 
before the air leak in port F8 occurred, with the vessel temperature between 
$80\,^\circ$C and $53\,^\circ$C. The offset of the linear fit is mainly attributed 
to the gauge on F9. The slope is used to estimate the activation of the NEG pumps. The fit in (b)  
is used to estimate the pressure that the spectrometer would have reached without the air leak.}
\label{fig:pressure_ratio} 
\end{figure}

With the vacuum sleeve around the leaky flange working, the Extractor gauges measured a pressure of
$3.5\cdot 10^{-11}\,$mbar in pump port P3, and $1.7\cdot 10^{-10}\,$mbar in port F9 at 
a temperature of $20\,^\circ$C. 
As mentioned before, the pressure measured in F9 had an offset that was caused by local 
outgassing in the order of  $10^{-10}\,$mbar. Fig.~\ref{fig:pressure_ratio}a shows a plot of the pressure in 
P3 against F9, measured just before the air leak opened up at port F8. During that period 
the temperature of the spectrometer dropped from $80\,^\circ$C to $53\,^\circ$C. The Extractor 
gauges, mounted outside the thermal insulation of the Main Spectrometer, were already close to 
the ambient temperature of $20\,^\circ$C inside the experimental hall. Thus no temperature 
correction was applied to the measured pressure. A linear fit to the pressure data in 
Fig.~\ref{fig:pressure_ratio}a provided an offset 
between the two gauges of $1.8\cdot 10^{-10}\,$mbar. Although the fit can only determine 
the difference between the offsets of both gauges, it was assumed that the offset can
be attributed to F9, since the pressure in P3 was already well below this value.  

If the offset, caused by other gas species, cannot be neglected, the hydrogen ratio 
between F9 and P3 cannot be determined directly from the ratio 
of the absolute pressure values, but only through the slope of a linear fit on data 
with varying hydrogen pressure and constant contributions from other gases. Due to a 
slightly varying leak rate in F8 and constant hydrogen outgassing at room temperature, 
this condition was only fulfilled before the air leak occurred.
The slope of 0.41 of the fit has been used to estimate the activation of the NEG pump.
 
\subsection{Activation of the NEG pumps \label{sec:NEGpump}}

The thermal activation of the NEG pumps at the recommended temperature of
$350\,^\circ$C for a duration of $24\,$hours was not possible. Reducing
either the temperature or the activation time can result in a lower pumping speed,
and/or a reduced capacity for gas. The actual activation during the bake-out
campaign lasted $28\,\mathrm{h}$ at $300\,^\circ$C. 

The simulations described in Sec.~\ref{sec:simulation_H2} were used to estimate the 
level of activation and the effective pumping speed of the NEG pumps.
Comparing the plot in Fig.~\ref{fig:simulation_NEG}.a to the fitted slope of 0.41 
(Fig.~\ref{fig:pressure_ratio}.a) of the measured data results in a sticking coefficient 
of 1.1\%. Assuming a sticking coefficient of 2.9\% for fully activated NEG strips \cite{lit:SAES}, 
baking at $300\,^\circ$C for $28\,$h led to an activation of 40\% of the nominal pumping speed of
the NEG strips. 
  
For the partly activated NEG pump with $\alpha = 1.1\%$, the effective pumping speed 
was $96\,$m$^3$/s (see Fig.~\ref{fig:simulation_NEG}.b). This is already 77\% of the 
maximum effective pumping speed. For three NEG pumps and six TMPs the effective 
pumping speed for hydrogen in the MS added up to $300\,$m$^3$/s. 

\begin{figure}
\centering
\includegraphics[width=0.9\textwidth]{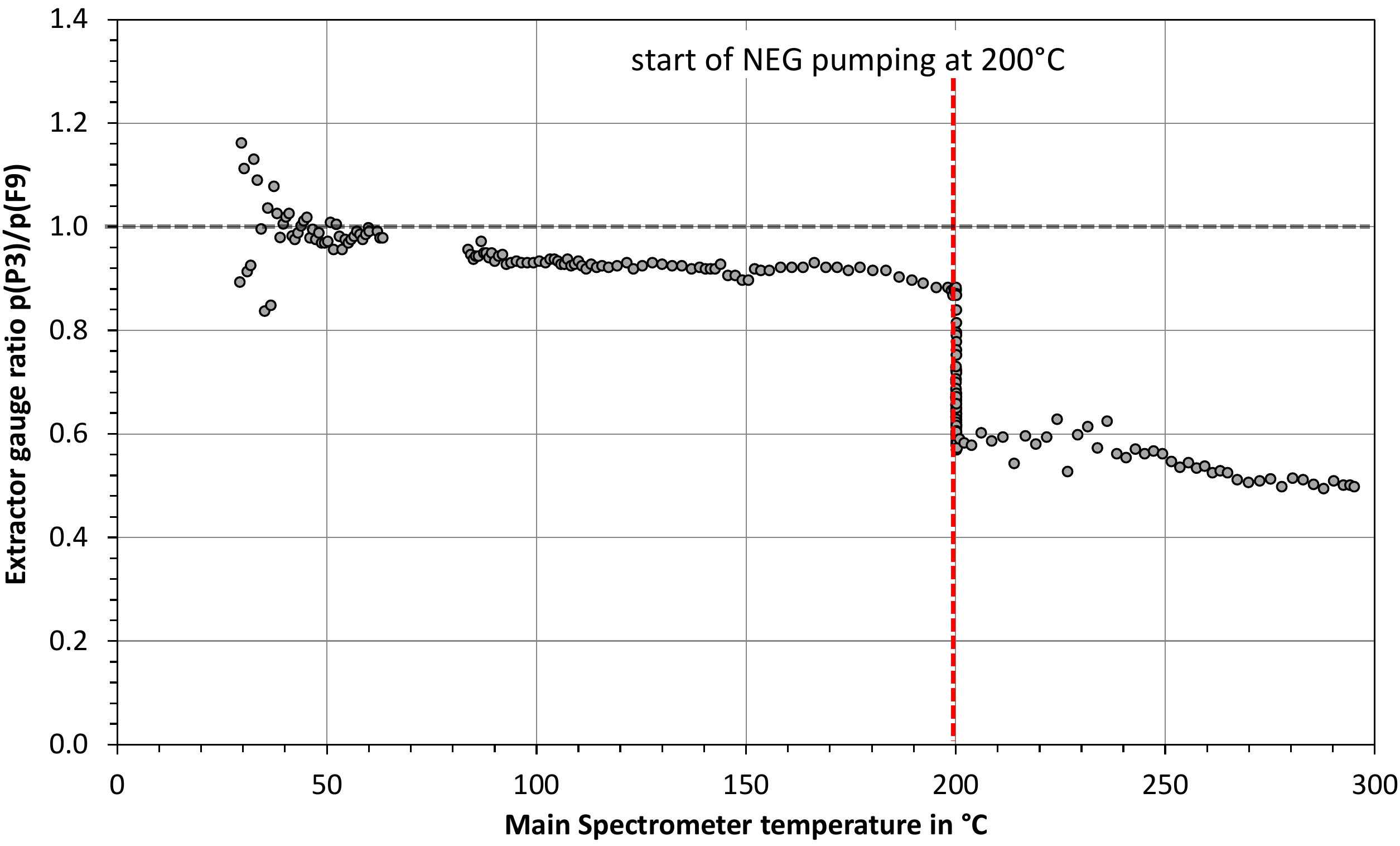} 
\caption[Start of NEG activation]{Ratio of the two Extractor
gauges $p_{F9}$ and $p_{P3}$. The start of the deviation of the pressure
ratio during the initial heating indicates the start of the NEG pumping. Apparently the NEG strips
started pumping already at a temperature of $200\,^\circ$C.}
\label{fig:NEGstart} 
\end{figure}

In general NEG pumps also pump nitrogen, water, CO, CO$_2$, and other active gases. 
While hydrogen is pumped by the reversible process of physisorption, other gases are 
pumped by irreversible chemisorption. During the activation process hydrogen is 
released and pumped out by TMPs, while compounds with other gas species on the surface 
diffuse into the bulk  of the NEG material, leaving a clean and reactive metal 
surface, ready to pump again \cite{lit:manini_neg2009}. During the initial heating, 
the ratio of pressures in Fig.~\ref{fig:NEGstart}, 
which was hydrogen dominated above $150\,^\circ$C, shows that the NEG pumps 
started pumping already at $200\,^\circ$C. The pressure ratio can be compared to the simulated 
ratio as a function of the sticking coefficient of the NEG strips in Fig.~\ref{fig:simulation_NEG}.
During the air leak the slope for the nitrogen-dominated gas composition was measured to be 0.94. 
This leads to the conclusion that the NEG pumps have either been saturated during the leak, 
or not been activated at all for gases other than hydrogen. Thus these gases were mainly 
pumped by the TMPs. 

\subsection{Estimation of the outgassing rate \label{sec:outgassing}}

In general the outgassing rate of a surface can be measured with a rise of pressure measurement,
where the vacuum vessel is evacuated to ultra-high vacuum before it is isolated 
from the pumps.
This method has been applied before on the Main Spectrometer, revealing a slowly decreasing hydrogen 
outgassing rate of the stainless steel from 
1.5 to $1.2\cdot 10^{-12}\,\rm{mbar}\cdot \ell/\rm{s\cdot cm}^2$ \cite{lit:JVS09}.  

However, with the activated 
NEG pumps inside the pump ports without an isolating valve, this method could not be used.
Therefore the outgassing rate $j_{\rm{H}_2}$ for hydrogen was estimated by 
multiplying the hydrogen pressure $p_{\rm{H}_2}$ in the main volume with the effective pumping 
speed $S_{\rm eff} = 300\,$m$^3$/s, normalized to the inner surface $A= 1271\,\rm{m}^2$:
\begin{equation}
j_{\rm{H}_2} = \frac{p_{\rm{H}_2}\cdot S_{\rm eff}}{A}. \label{equ:outgassing}
\end{equation}
The outgassing area includes all surfaces at $20\,^\circ$C, and excludes only the activated getter surface.
With the outgassing rate depending strongly on the temperature, the pressure had to be determined at
the standard operating temperature of $20\,^\circ$C. Since the Extractor gauge at port F9 had a non-negligible 
offset, which decreased over time and could not be determined accurately enough at 20$\,^\circ$C,
the temperature-dependent, hydrogen-dominated pressure, measured before the air leak, 
was extrapolated to $20\,^\circ$C.  
Atoms and molecules on a surface oscillate rapidly at a frequency $\nu_0 \approx 10^{13}\,$Hz. If their kinetic 
energy is above the desorption energy $E_{\rm{des}}$, they can escape the surface into the volume 
of the spectrometer. With $N$ particles on a surface area $A$, $\Delta N = N\cdot \rm{exp}(-E_{\rm{des}}/R\cdot
 T_{\rm{W}})$ particles meet this requirement for a wall temperature 
$T_{\rm{W}}$ \cite{lit:jousten}. $R$ is the molar gas constant. The surface desorption rate, 
which is equivalent to the outgassing rate, is
\begin{equation}
j_{\rm{des}} = \frac{1}{A}\cdot \frac{\rm{d}N}{\rm{d}t} = -\nu_0\cdot \frac{N}{A}\cdot 
e^{-E_{\rm{des}}/R\cdot T_{\rm{W}}}.  \label{equ:desorption-rate}
\end{equation}
According to Equ.~\ref{equ:pumping-speed} the effective pumping speed of the NEG pumps 
is proportional to $\bar{c}$, which in turn is proportional to the square root of the gas temperature $T_{\rm{W}}$.
Assuming that the sticking coefficient $\alpha_{\rm{NEG}}$, and thus the pumping probability $w$,
does not change much in the temperature range of the measurement, the effective pumping speed at
temperature $T_{\rm{W}}$ is
\begin{equation}
S_{\rm{eff}}(T_{\rm{W}}) = S_{\rm{eff}}(293\,\rm{K})\cdot \frac{\sqrt{T_{\rm{W}}}}{\sqrt{293\,\rm{K}}}. 
\label{equ:S_eff}
\end{equation}
The pressure $p_{\rm{m}}$ measured by an Extractor gauge is proportional to the particle density 
\begin{equation}
n = \frac{p}{k_{\rm{B}}\cdot T} = \frac{j_{\rm{des}}\cdot A}{S_{\rm{eff}}(T_{\rm{W}})}. 
\label{equ:density}
\end{equation}
With Eqs.~\ref{equ:outgassing}, \ref{equ:desorption-rate}, and \ref{equ:density}, the measured pressure
 as a function of the wall temperature can be expressed as
\begin{equation}
p_{\rm{m}} = a_0\cdot \frac{\sqrt{293\,\rm{K}}}{\sqrt{T_{\rm{W}}}} \cdot e^{a_1\cdot \frac{1}{T_{\rm{W}}}}
\label{equ:p_m}
\end{equation}
A fit of the data from the Extractor gauge (F9) in Fig.~\ref{fig:pressure_ratio}.b provided 
parameters $a_0$ and $a_1$, which were used to extrapolate the pressure in the main volume at $20\,^\circ$C.
The fitted value of $2.6\cdot 10^{-11}\,$mbar has to be multiplied with the 
gas correction factor of gauge F9 for hydrogen (2.2), in order to get the real hydrogen pressure of
$p$($20\,^\circ$C) = $5.7\cdot 10^{-11}\,$mbar inside the Main Spectrometer.  Inserting all numbers in
Equ.~\ref{equ:outgassing}, we derive an outgassing rate at 
$20\,^\circ$C of $j_{\rm{H}_2} = 1.4\cdot 10^{-12}\,\rm{mbar}\cdot \ell/\rm{s\cdot cm}^2$. 

The partial pressure, measured with the RGA about two months after the bake-out
(see Fig.~\ref{fig:rga_baking}), was $1.9\cdot 10^{-11}\,$mbar. 
Since the RGA data have been calibrated against Extractor gauge P3, it has to be multiplied with a gas 
correction factor of 2.3 and with the pressure ratio (1/0.41) between hydrogen in the main volume and in 
pump port 3, the location of the RGA. The real hydrogen pressure in the main volume, determined with 
the RGA, is p($20\,^\circ$C) = $1.1\cdot 10^{-10}\,$mbar, corresponding to an outgassing rate of 
$j_{\rm{H}_2} = 2.5\cdot 10^{-12}\,\rm{mbar}\cdot \ell/\rm{s\cdot cm}^2$. 
While the uncertainty of the Extractor
calibration is 10\% at $10^{-6}\,$mbar (used here as a reference) the main uncertainties of the 
pressure cannot be quantified due to the linear extrapolation over almost five orders 
of magnitude.  
 
The extrapolation from higher temperatures to $20\,^\circ$C and the measurement with the 
RGA find within uncertainties comparable results for the H$_2$ partial
pressure, and therefore, for the outgassing rate. The outgassing rate is in the same range as that in the 
first measurements in 2007 \cite{lit:JVS09}, where the inner electrode system and the NEG pumps 
were not yet installed and the final bake-out temperature reached the nominal value of  $350\,^\circ$C.
This result also shows that the vacuum quality was not affected after five years of inner electrode 
system installation and other construction work under cleanroom condition in the vessel.


\section{Preparations for first spectrometer measurements\label{sec:argon}}

\begin{figure} [th]
\centering
\includegraphics[width=\textwidth]{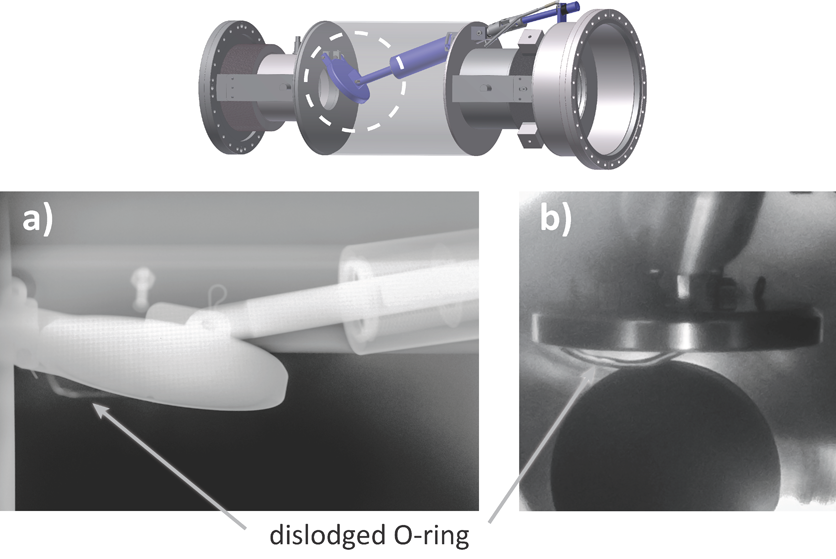} 
\caption[Beam-line valve]{A partly dislodged O-ring (a: X-ray image, b: photo taken during valve
repair) prevented the inline valve between spectrometer and
detector from being closed. }\label{fig:inline_valve} 
\end{figure}

The last step of the commissioning, before background and transmission measurements with electrons 
could start, was to connect  the electron source at the source end of the spectrometer 
and the detector system at the opposite end.
Both components had been decoupled from the spectrometer during the bake-out process, in order to protect the 
ceramic beam-line insulators from mechanical forces due to thermal expansion (up to 12$\,$cm) and contraction 
during temperature cycling. During bake-out the inline valves (see Fig.~\ref{fig:inline_valve}.c), 
connecting the insulators with detector 
and electron source, were closed with blank flanges, and the flaps of both valves were in the open position. 

\subsection{Locating a leak in the beam-line valve \label{sec:valve}}

After the bake-out was finished, both valves were to be closed and the outer section vented with grade 6.0 
argon (contamination with other gas species $\sim 10^{-6}$)
before the blank flanges were removed for connections to external components. 
The valve at the source end performed as
expected. When the detector valve was slowly vented with argon, the pressure in the spectrometer rose 
immediately, indicating that the valve could not be closed properly. As long as the blank flange was in place this
serious leak posed no threat for the vacuum of the spectrometer. However, the detector could not be connected.
Venting of the spectrometer with air was not an option, since it would deactivate the NEG pumps. 

Before finding a remedy, the problem had to be diagnosed. 
The whole valve was X-rayed from several directions and for several positions of the flap. The result is 
shown in Fig.~\ref{fig:inline_valve}.a. One can clearly recognize the 
Kalrez\textsuperscript{\textregistered} O-ring, which had slipped out of its 
groove. After identifying the cause of the leak, a method had to be devised to vent the spectrometer, open
the blank flange of the valve, and replace the O-ring without deactivating the NEG pumps.


\subsection{Argon venting of the spectrometer \label{sec:venting}}

\begin{figure} 
\centering
\includegraphics[width=\textwidth]{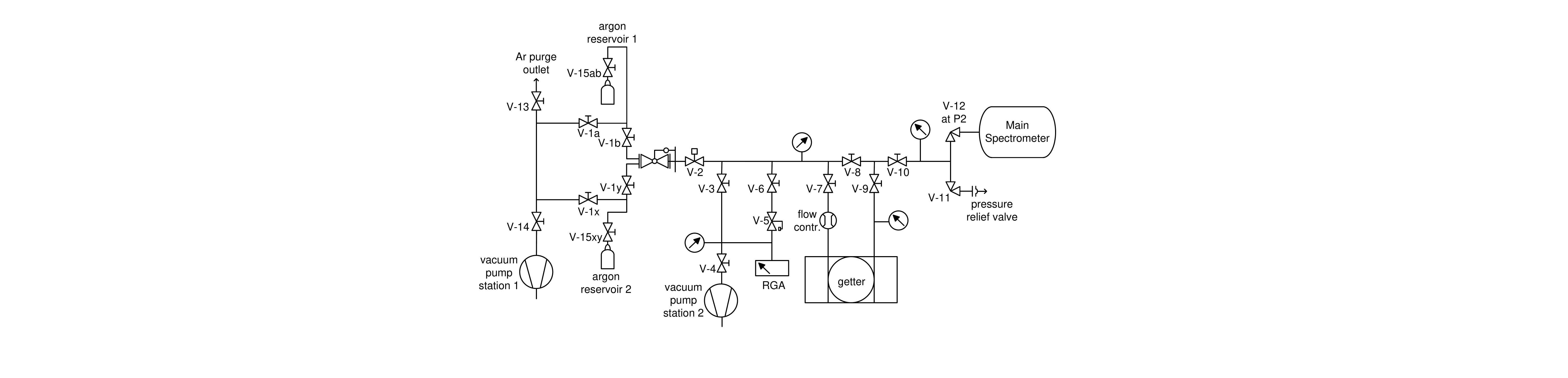}
\caption[Venting scheme]{Flowchart of the venting system, used for 
venting the KATRIN Main Spectrometer with ultra-clean argon. } 
\label{fig:venting_scheme} 
\end{figure}

Calculations showed that $1240\,\rm{m}^3$ grade-6.0 argon at atmospheric pressure, the best off-the-shelf argon 
available in large quantity, still contained too much nitrogen, oxygen, argon and water for the 
NEG pump activation to survive. Therefore a venting system was designed to provide ultra-clean argon 
of grade 9.0 (contaminations $\sim 10^{-9}$). The ultra-clean argon was produced while 
venting, by purifying commercially available grade 6.0 argon with a gas purifier system.
The key component of this system was the hot getter unit SAES PS4-MT50-R that uses a 
hot, zirconium-based getter cartridge \cite{lit:gas-purifier}. The gas purification system was provided 
on short notice by the M\"unster University group of the XENON collaboration.
The venting schematic is shown in Fig.~\ref{fig:venting_scheme}.

The hot getter works like a continuously activated NEG pump. In particular, gas species 
that are pumped by physisorption, such as nitrogen and oxygen, would cover the 
NEG getter surface rapidly and reduce its pumping speed. In hot getter material these 
compounds diffuse quickly into
the bulk of the getter, freeing the surface for continuous pumping.  
The purifier system is designed for high flow rates up to 100$\,$slpm 
(standard liters per minute) at a minimum inlet pressure of 2.8$\,$bar. 
For an inlet gas purity of 99.9995\% (which is fulfilled for the argon 6.0) 
the system is capable of purifying it to a purity of better than 1 ppb (part per billion) 
per contaminant species.

The grade-6.0 argon was delivered in bundles of $12 \times 50\,\ell$ bottles at 220$\,$bar. 
During the venting process 11 bundles were used for the spectrometer to reach atmospheric pressure.
There were always two bundles connected to the system, with only one bundle opened to the spectrometer.
When it reached a pressure below 20$\,$bar, the
other bundle was opened, which allowed a continuous venting process without interruption. 

The feed lines to both bundles had a connection to a scroll pump
and to a blow-off valve (via valves V-1x and V-1a in Fig.~\ref{fig:venting_scheme}), 
which were used to remove air from the pipes after
connecting a new bundle of bottles to the system. 
Only after the section with the new bundle was flushed with argon, 
with the blow-off valve opened, evacuated, and filled with clean argon again, was it ready to be connected 
to the purifier by opening the valve V-1b or V-1y, respectively.
The pressure regulators of the bundle were set to 4.5 - 5 bar.

Before starting the venting, the entire system was evacuated using pump station
2 (TMP and diaphragm pump), and flushed several times with clean argon, in order to
remove all traces of air from the gas lines. All connections on the low pressure
side were either made with VCR and CF flanges, or by orbital-welding
of stainless steel tubes. After setting up the system, it was leak-tested with a 
sensitivity of  $< 10^{-9}\, \rm{mbar\cdot \ell/s}$.

\begin{figure}
\centering
\includegraphics[width=0.9\textwidth]{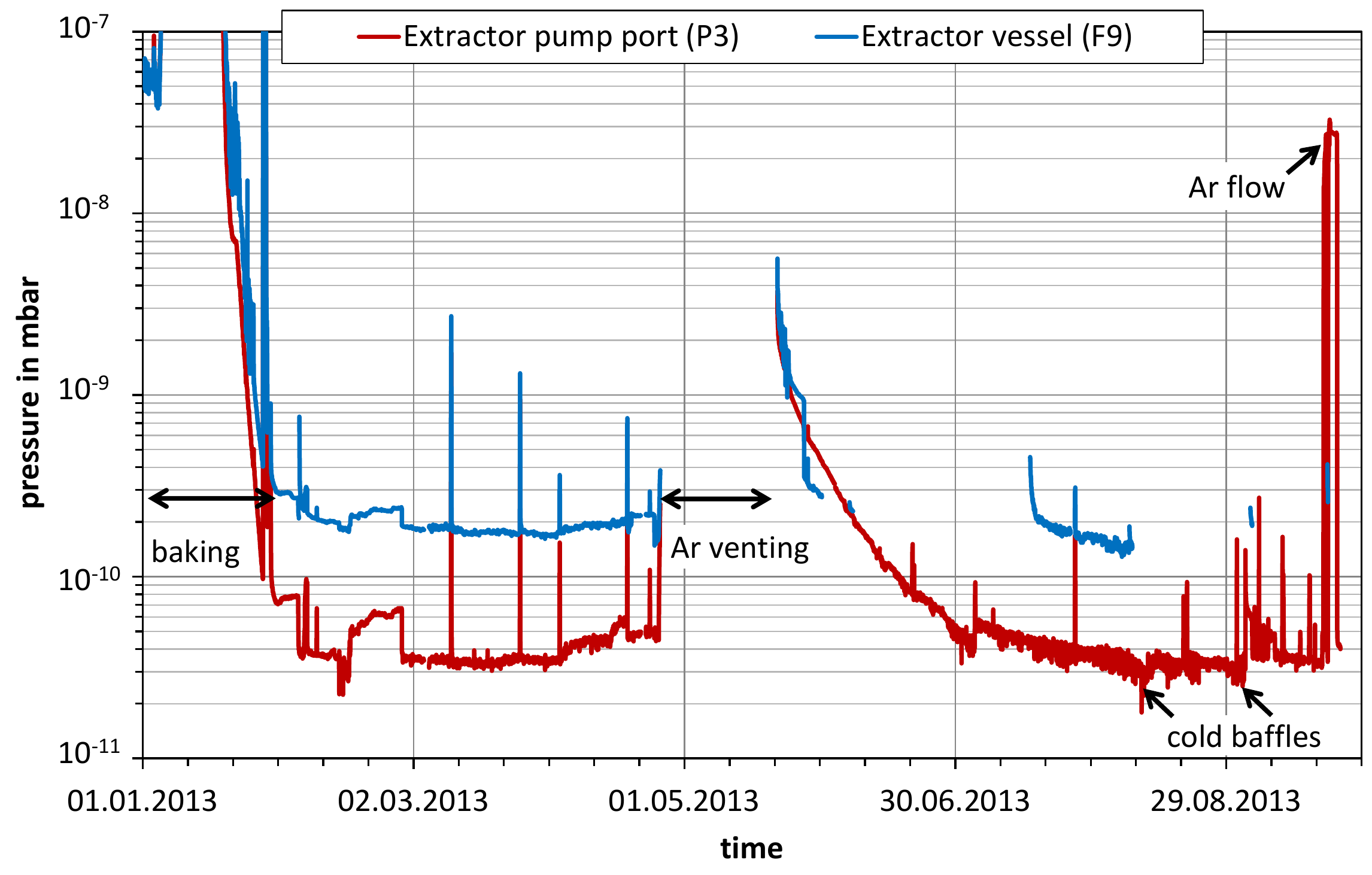}
\caption[Pressure in MS and pump port]{Overview of the progression of pressure in the 
main volume and in pump port 3, measured with the Extractor gauges. 
}\label{fig:pressure_all} 
\end{figure}

A quadrupole mass spectrometer was used to monitor for gross impurities
in the argon gas before being purified by the getter. Due to the high pressure in the feed line 
of the purifier, the RGA was operated behind a leak valve that is pumped by a TMP. The resolution of the RGA was 
not sufficient to detect impurities at the ppm level, but it was able to detect air leaks when new 
argon bottles were connected to the system. 
The amount of argon flowed into the system was measured by a MKS 1579A\textsuperscript{\textregistered} 
mass flow controller. The pressures at the inlet and the outlet of the getter cartridge were monitored by Swagelok PTU-S-AC9-31AD\textsuperscript{\textregistered}
capacitance pressure sensors. For safety reasons a pressure relief valve with an opening 
pressure of 0.2$\,$bar was installed to protect the spectrometer from overpressure. 

After the spectrometer was filled to atmospheric pressure, a polyethylene plastic bag was attached to
the end of the beam-line valve, and enclosed the blank flange that had to be opened for repairing the valve.
Clean tools and a replacement O-ring had been placed inside before attaching the bag to the valve.
The air tight bag had two gloves incorporated, thus serving as a ``flexible glove box''. It 
was evacuated and flushed with clean argon several times before carefully opening the blank flange.
After replacing the O-ring and closing the blank flange the spectrometer was evacuated again. 
The whole procedure, from the start of venting back to UHV, took 24 days. 

With the valve repaired, the detector system was connected, and everything 
was ready for the start of the electron measurements.

\subsection{Vacuum performance during spectrometer measurements
\label{sec:performance}}

\begin{figure}
\centering
\includegraphics[width=0.9\textwidth]{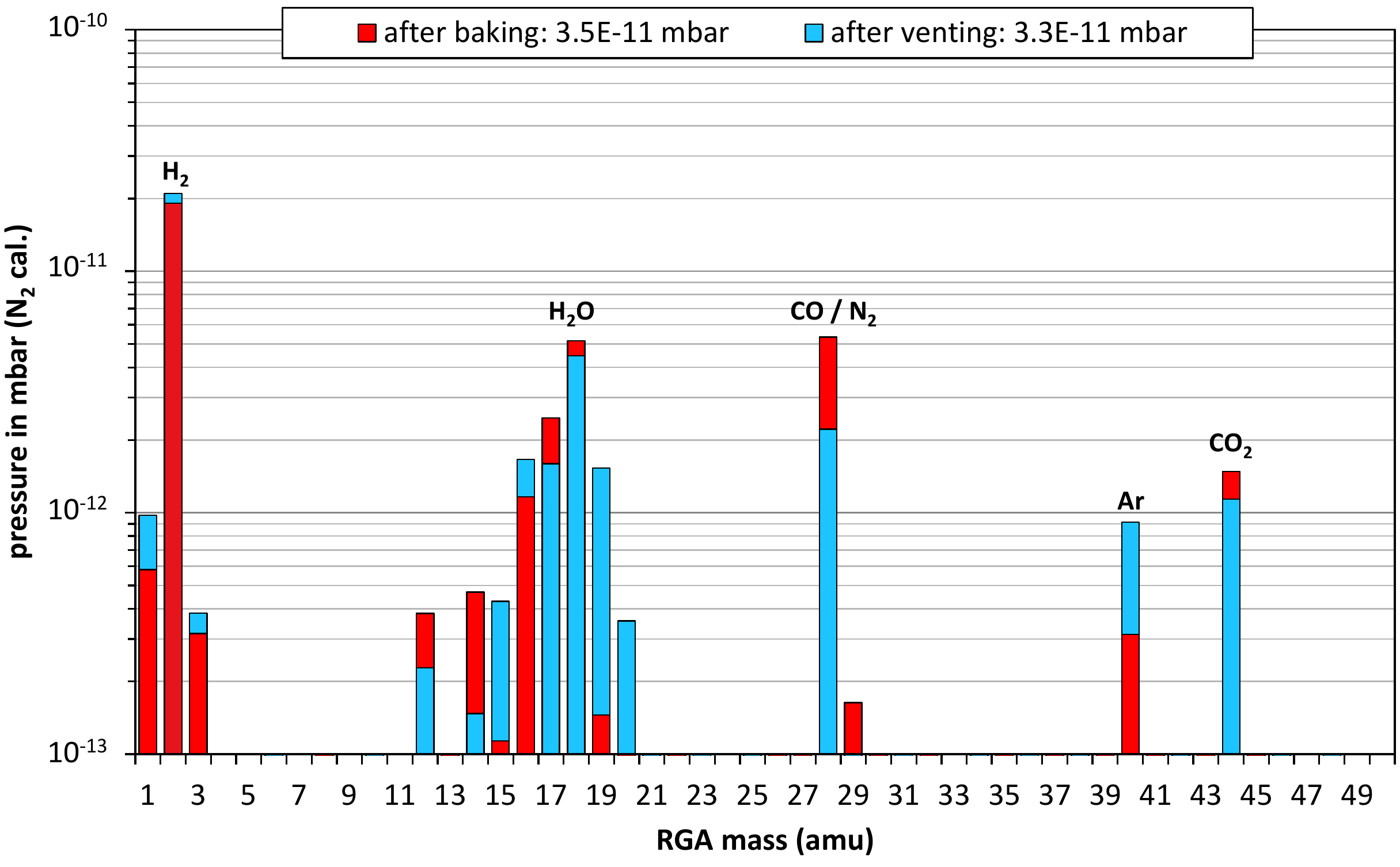}
\caption[RGA spectrum before and after venting]{RGA spectra before and after
venting with ultra-clean argon. After venting the hydrogen signal (mass 2) went down by a
factor of 30, while most other gases were only slightly reduced. Since the absolute pressure,
measured with the Extractor gauge, remained stable, it was assumed that the sensitivity of the RGA changed 
after venting with argon.
}\label{fig:RGA_venting} 
\end{figure}

During pump-down another air leak opened up at a CF flange at port F9. Like the 
leak at port F8, the flange connection was enclosed in a vacuum sleeve subsequently and pumped differentially.
Although the remaining leak rate was low enough for the spectrometer measurements, there were 
additional fluctuations in the pressure offset of the Extractor gauge due to its close proximity to the leak.
Therefore the Extractor gauge was switched off for most of the time.

Below $10^{-8}\,$mbar the pressure dropped only slowly. It took about two months
until the spectrometer reached the same low pressure it had before venting 
(see Fig.~\ref{fig:pressure_all}). For most of the time the pressure was dominated by argon.
However,  the pressure was sufficient for the early measurements, and the slow
desorption rate of argon did not delay the schedule. 
Sharp changes in the pressure, seen in Fig.~\ref{fig:pressure_all}, were mainly caused by 
maintenance at  the differentially pumped leaks, opening and closing of the valve to the electron source, 
short tests of the cryogenic baffles, and measurements at elevated pressure (argon: $3\cdot 10^{-8}\,$mbar)
for detailed investigations of the radon-related background rate \cite{lit:phd:goerhardt}.  

After reaching the low pressure regime again, the Extractor gauge in pump port 3 measured a pressure of
$3.3\cdot 10^{-11}\,$mbar. The Extractor gauge in the main volume on port F9 
read a value of $1.4\cdot 10^{-10}\,$mbar, slightly lower than before the venting. However, the uncorrected 
mass spectrum displayed a more dramatic effect. While the signals from
most gas species were of similar size as before the venting, the hydrogen peaks (mass 1, 2 and 3) 
dropped by a factor of 30. 
For both measurements the same RGA settings were used. As described in appendix~\ref{sec:RGAcalibration}, 
it was assumed that the RGA's sensitivity for hydrogen changed during the venting, and a new calibration factor was 
determined. The RGA spectra before and after venting are shown in Fig.~\ref{fig:RGA_venting}.
Despite venting the spectrometer to full atmospheric pressure with ultra-clean argon, the (corrected)
RGA peaks, as well as the absolute pressure, remained basically the same, which in 
turn leads to the conclusion that the NEG pumps were still active after pump-down. 

The absolute pressure in the main volume can be estimated with the pressure of the
Extractor gauge in pump port 3 ($3.3\cdot 10^{-11}\,$mbar), and the hydrogen partial pressure of the 
RGA ($2.1\cdot 10^{-11}\,$mbar). The difference of
$1.2\cdot 10^{-11}\,$mbar is attributed to the remaining gas species. Simulations in Sec.~\ref{sec:simulation}
showed that the pressures in the main volume and in the pump ports are approximately the same,
if the pumping speed is small compared to the conductance of the baffles. With the pressure ratio of
1.06 observed during the air leak, it is assumed that this condition is fulfilled for all gas species 
but hydrogen. Adding the remaining pressure and the hydrogen partial pressure, 
applying the correction for the pressure drop at
the baffle (1/0.41) and multiplying this corrected value by the gas correction factor (2.3) results in an 
absolute pressure in the main volume of $1.3\cdot 10^{-10}\,$mbar. This is very close to the pressure 
of $1.4\cdot 10^{-10}\,$mbar that was measured by the Extractor gauge at port F9.


\section{Conclusions \label{sec:conclusions}}

In this work we have described the vacuum system of the 23.2$\,$m long 
Main Spectrometer of the KATRIN experiment and reported on the details of its successful 
commissioning. The simulated nominal pumping speed for hydrogen of 
the three NEG pumps amounts to almost 1,000\,m$^3$/s. It has been designed to reach ultra-high vacuum 
in the range of $10^{-11}\,$mbar, for an expected outgassing rate at room temperature of 
$10^{-12}\,\rm{mbar}\cdot \ell/\rm{s\cdot cm}^2$. The effective pumping speed, 
reduced by the cryogenic baffles 
required for the reduction of radon-related backgrounds, adds up to 375\,m$^3$/s, 
if the getter were activated at 350$\,^\circ$C for 24$\,$h. The actual effective pumping speed was 
300\,m$^3$/s, which was reached after activating a total of $3,000\,$m of SAES 
St707\textsuperscript{\textregistered}  NEG strips at  
300$\,^\circ$C for 28$\,$h. With a value in the range of 
$1.4\,-\,2.5\cdot 10^{-12}\,\rm{mbar}\cdot \ell/\rm{s\cdot cm}^2$ the 
estimated hydrogen outgassing rate of the stainless steel walls was already close to the expected value. 
The total absolute pressure in the main volume, which was reduced by more than three orders of magnitude 
by the baking of the MS, reached a value of around $10^{-10}\,$mbar. The residual gas composition was 
dominated by hydrogen, which made up about 90\% of the total pressure.  The rest was mainly composed 
of water, CO and CO$_2$.

The lessons learned from the problems that occurred during the commissioning 
measurements led to several modifications in the design of the vacuum system, 
which have been implemented and tested during the two shutdown periods in 2014 and 2015, 
followed by pump-downs and commissioning measurements with and without baking at $200\,^\circ$C:
\begin{itemize}
\item Some of the problems with the mechanical stability of the CuBe high-voltage wires that led to 
electrical short circuits between the inner and outer wire layers of the modules of the inner electrode 
system during bake-out have been solved. About half of the electrode system is currently free of 
short circuits. The difficult and time-consuming repair of the remaining short circuits has been postponed 
since recent electron background measurements revealed that the present MS background rate would not 
be significantly reduced by a full dual layer operation of the wire electrode system. 
\item A redesign of the NEG pumps for electrical heating allows a vessel bake-out at lower temperatures. 
In the original design, 
the NEG strips were heated by radiation from the hot spectrometer walls. In the new design the temperature of the 
spectrometer can be as low as 200$\,^\circ$C during the local activation process at 400$\,^\circ$C. 
This measure provides an additional safety margin for the wires of the electrode system.
\item After several leaks occurred at CF flanges with standard gaskets (2-mm thick), 
they have been replaced by thicker,
3$\,$mm copper gaskets, resulting in a larger travel for re-tightening of the bolts, if needed.  
So far no further leaks have occurred.
\item The Extractor gauge at port F9 has been moved from the 40$\,$mm tube at the side of the port
to the top, where it looks directly into the main volume through a 100$\,$mm adapter and valve. 
The lowest base pressure measured after the bake-out and activation of one NEG pump was 
$6\cdot 10^{-11}\,$mbar (nitrogen calibration), compared to $1.7\cdot 10^{-10}\,$mbar 
with the old design and three activated NEG pumps. 
\item At port F10 a calibrated orifice and a Baratron\textsuperscript{\textregistered} gauge 
have been added in front of a leak valve. This
allows more accurate flux measurements with different gases for in-situ calibrations of the gauges and a
more accurate determination of the effective pumping speed. 
\item The groove for the Kalrez\textsuperscript{\textregistered} O-ring in the flapper of the 
in-beam valve has been redesigned to prevent the displacement of the seal during bake-out.
\end{itemize}

\noindent A very valuable lesson learned from the mishap with the Kalrez\textsuperscript{\textregistered} 
O-ring was that for smaller repairs we can vent the spectrometer to atmospheric pressure with ultra-clean argon, 
without deactivating the NEG pumps. Grade 6.0 argon, the best quality of bottled argon available, was further 
cleaned by a hot NEG-based gas purifier that reduced the impurities by another three orders of magnitude.
With this method the NEG pumps retained their initial pumping speed, and the absolute pressure before and 
after venting was virtually the same. 

In the last commissioning measurements in 2014 and 2015, the spectrometer was operated with 
only one activated NEG pump, successfully testing the new electrical heating concept. 
At around $10^{-10}\,$mbar the pressure dependence of the background rate was negligible. 
Therefore it has been decided to operate the spectrometer with only two NEG pumps for 
the next measurements. Thus, enough of the 
special low-activity NEG strips remain as spares to replace at least one NEG-pump, if necessary.
A new, high voltage insulated liquid nitrogen feed-line for the cryogenic baffles has also been installed, 
demonstrating the reliable suppression of radon background from the NEG pumps.

For the first tritium measurements, following the final engineering runs after merging the 
\emph{Source and Transport Section} with the \emph{Spectrometer and Detector Section} in 2016, 
we expect for two electrically activated NEG pumps an absolute pressure below $1\cdot 10^{-10}\,$mbar, 
dominated by hydrogen. Final results for the effective neutrino mass are expected five years after 
starting the tritium measurements. 


\acknowledgments
We want to thank Christian Day, Volker Hauer and Xueli Luo from the Institute for Technical Physics at KIT,
as well as the ASTEC vacuum group at Daresbury lab (Joe Herbert, Oleg Malyshev, Keith Middleman, and Ron Reid) 
for many helpful discussions and their contributions to the design of the main spectrometer vacuum system.
In addition we want to thank Volker Hauer for calibrating our vacuum gauges with his calibration system.
We also thank our colleagues from the XENON group at M\"unster University for providing the gas purification system,
which was vital for the successful commissioning of the Main Spectrometer.
We acknowledge the support of the German Helmholtz Association (HGF), 
the German Ministry for Education and Research BMBF (05A14VK2 and 05A14PMA),
the Helmholtz Alliance for Astroparticle Physics (HAP), 
the Grant Agency of the Czech Republic (GACR) P203/12/1896,
and the US Department of Energy through grants DE-FG02-97ER41020 , DE-FG02-94ER40818, 
DE-SC0004036, DE-FG02-97ER41041, and DE-FG02-97ER41033. 
Lawrence Berkeley National Laboratory (LBNL) is operated by The Regents of the University of 
California (UC) for the U.S. Department of Energy (DOE) under Federal Prime Agreement DE-AC02-05CH11231.


\begin{appendix}
\section{Estimation of the RGA calibration constants \label{sec:RGAcalibration}}

The external calibration of the RGA for different gas species at pressures between $10^{-7}$ 
and $10^{-6}\,$mbar for the SEM-detector showed a non-linear behavior. Therefore this calibration method
was not suitable for  linear extrapolation over a range of 5 orders of magnitude, down to 
the $10^{-11}\,$mbar pressure regime.

Since the tight measurement schedule did not allow for detailed in-situ calibration measurements,
the RGA peaks (SEM detector) had to be roughly calibrated against the nitrogen-calibrated signal of 
Extractor gauge P3 using existing data. This analysis was done for three gas species: 
hydrogen (mass 2), argon (mass 40, 36, and 20) and nitrogen (28, 14). 
For all other mass peaks the nitrogen calibration was used. The nitrogen and argon calibrations were determined
by comparing the pressure changes of the Extractor gauge, and the appropriate RGA peaks at several occasions 
when the partial pressure of the respective gas species changed. The results before and after the venting with argon 
were in good agreement, implying a stable SEM gain for these gas species.  
The argon calibration factor was 1.5 times larger than the value for nitrogen.
This number is close to the inverse of the argon gas correction factor of the Extractor gauges (0.7), suggesting
that the sensitivities of the RGA for nitrogen and argon are almost the same. 

Since the hydrogen pressure was stable for most of the time at 20$\,^\circ$C, the calibration factor 
was determined by adjusting it to the difference between the absolute pressure of the Extractor gauge 
and the sum of the other calibrated RGA peaks, excluding mass 2.  If the gas correction factors for the 
Extractor gauge and for the RGA in SEM mode were the same, one would expect the same correction 
factor as for nitrogen. However, three different time intervals were identified where the hydrogen 
calibration factor changed dramatically. The first interval was during the bake-out period with a correction factor
of 0.34 times the nitrogen factor, thus indicating a sensitivity that is three times higher for hydrogen. The second interval
started after a large air leak opened up (see Section~\ref{sec:baking}). The correction factor changed to
0.086 times that of nitrogen. The third interval started after the spectrometer was vented with ultra-clean 
argon to atmospheric pressure (see Section~\ref{sec:venting}). The correction factor changed 
to a value of 2.6 times that of nitrogen, which would imply a decrease of the hydrogen sensitivity 
of the SEM detector by a factor of 30. Within each time interval the hydrogen signal remained stable. It is not clear
why the hydrogen sensitivity of the RGA would change so dramatically, in particular after the argon venting.
However, assuming that the hydrogen sensitivity was the same before and after the venting, it would 
imply that the hydrogen outgassing of the stainless steel has decreased by a factor of 30. With 
basically the same absolute pressure measured by the Extractor gauge before and after the venting,
this assumption seems very unlikely compared to a changing sensitivity for mass 2.  Therefore we applied
the different calibration factors for the mass-2 peaks at different time intervals.

Since the RGA peaks were calibrated against the 
nitrogen calibration of the Extractor gauge, one has to apply the gas correction factors of the 
Extractor gauge, if the real partial pressure needs to be determined. 

\end{appendix}

\clearpage 


 \end{document}